\documentclass[preprint,12pt]{elsarticle}

\usepackage[T1]{fontenc}
\usepackage{hyperref}
\usepackage{url}
\usepackage{amsfonts}
\usepackage{lipsum}
\usepackage{booktabs}       % professional-quality tables
\usepackage{graphicx}
\usepackage{caption}
\usepackage{comment}
\usepackage{mathtools}
\graphicspath{{media/}}
\usepackage[dvipsnames]{xcolor}
\usepackage{lineno}
\usepackage[export]{adjustbox}
\usepackage{upgreek}
\usepackage{xspace}
\xspaceaddexceptions{]}
\usepackage{amssymb}
\usepackage{subcaption}
\usepackage{footmisc}

\usepackage[table]{xcolor}

\definecolor{thick45colour}{HTML}{FF0000}
\definecolor{thick30colour}{HTML}{FFA500}
\definecolor{thick20chblcolour}{HTML}{32CD32}
\definecolor{thick20colour}{HTML}{008000}
\definecolor{thick35colour}{HTML}{0000FF}
\definecolor{thick25colour}{HTML}{79F6FC}

\journal{Nuclear Instruments and Methods in Physics Research A}

\date{1st July 2025}

\newcommand{\mevneut}{\ensuremath{\textrm{n}_{\textrm{eq}} \textrm{cm}^{-2}}\xspace}
\newcommand{\invmevneut}{\ensuremath{\textrm{cm}^{2} \textrm{n}_{\textrm{eq}}^{-1}}\xspace}
\newcommand{\micron}{\ensuremath{\rm{\upmu}\textrm{m}}\xspace}
\newcommand{\betanoitalics}{\ensuremath{\rm{\upbeta}}\xspace}
\newcommand{\atcm}{\ensuremath{\textrm{cm}^{-3}}\xspace}
\newcommand{\trdutmcp}{\ensuremath{\mathrm{\sigma}_{\textit{t}}(\rm{DUT; MCP})}\xspace}
\newcommand{\trdutmcpsquared}{\ensuremath{\mathrm{\sigma}^{2}_{\textit{t}}(\rm{DUT; MCP})}\xspace}
\newcommand{\trdutmcpsquaredone}{\ensuremath{\mathrm{\sigma}^{2}_{\textit{t}}(\rm{DUT1; MCP})}\xspace}
\newcommand{\trdutmcpsquaredtwo}{\ensuremath{\mathrm{\sigma}^{2}_{\textit{t}}(\rm{DUT2; MCP})}\xspace}
\newcommand{\trdutdutsquared}{\ensuremath{\mathrm{\sigma}^{2}_{\textit{t}}(\rm{DUT1; DUT2})}\xspace}
\newcommand{\trmcp}{\ensuremath{\mathrm{\sigma}_{\textit{t}}(\rm{MCP})}\xspace}
\newcommand{\trmcpsquared}{\ensuremath{\mathrm{\sigma}^{2}_{\textit{t}}(\rm{MCP})}\xspace}
\newcommand{\trdut}{\ensuremath{\mathrm{\sigma}_{\textit{t}}(\rm{DUT})}\xspace}
\newcommand{\trdutmin}{\ensuremath{\mathrm{\sigma}_{\textit{t}}(\rm{DUT})_{\textrm{min}}}\xspace}
\newcommand{\trdutsquaredone}{\ensuremath{\mathrm{\sigma}_{\textit{t}}^{2}(\rm{DUT1})}\xspace}
\newcommand{\trdutsquredtwo}{\ensuremath{\mathrm{\sigma}_{\textit{t}}^{2}(\rm{DUT2})}\xspace}
\newcommand{\trdutdut}{\ensuremath{\mathrm{\sigma}_{\textit{t}}(\rm{DUT1; DUT2})}\xspace}

\newcommand{\jitter}{\ensuremath{\mathrm{\sigma}_{\textrm{jitter}}}\xspace}
\newcommand{\landau}{\ensuremath{\mathrm{\sigma}_{\textrm{ionisation}}}\xspace}
\newcommand{\ps}{\ensuremath{\textrm{ps}}\xspace}
\newcommand{\fC}{\ensuremath{\textrm{fC}}\xspace}
\newcommand{\bulkconc}{\ensuremath{p_{\mathrm{bulk}}}\xspace}
\newcommand{\bulkconcunit}{\ensuremath{\textrm{cm}^{-3}}\xspace}
\newcommand{\bw}{\ensuremath{BW_{\mathrm{tot}}}\xspace}
\newcommand{\risetime}{\ensuremath{t_{\mathrm{rise}}}\xspace}
\newcommand{\efieldunit}{\ensuremath{\textrm{V}\rm{\upmu}\textrm{m}^{-1}}\xspace}
\newcommand{\tcfddutone}{\ensuremath{t_{\textrm{CFD}}(\textrm{DUT1})}\xspace}
\newcommand{\tcfdduttwo}{\ensuremath{t_{\textrm{CFD}}(\textrm{DUT2})}\xspace}

\begin{document}

\begin{frontmatter}

\title{Timing resolution from beam tests on thin LGADs down to 16.6 ps}

\author[1]{Robert Stephen White}
\author[1]{Marco Ferrero}
\author[2,1]{Valentina Sola}
\author[2,1]{Anna Rita Altamura}
\author[3,1]{Roberta Arcidiacono}
\author[5,4]{Maurizio Boscardin}
\author[1]{Nicol\`o Cartiglia}
\author[5,4]{Matteo Centis Vignali}
\author[6]{Tommaso Croci}
\author[2]{Matteo Durando}
\author[2]{Simone Galletto}
\author[6]{Alessandro Fondacci}
\author[2,1]{Leonardo Lanteri}
\author[2,1]{Ludovico Massaccesi}
\author[3,1]{Luca Menzio\fnref{affiliation_lmenzio}}
\author[6]{Arianna Morozzi}
\author[7,6]{Francesco Moscatelli}
\author[8,6]{Daniele Passeri}
\author[5,4]{Giovanni Paternoster}
\author[1]{Federico Siviero}

\affiliation[1]{organization={Istituto Nazionale di Fisica Nucleare, Sezione di Torino}, addressline={via P. Giuria 1}, city={Torino}, country={Italy}}
\affiliation[2]{organization={Universit\`{a} degli Studi di Torino}, addressline={via P. Giuria 1}, city={Torino}, country={Italy}}
\affiliation[3]{organization={Universit\`{a} del Piemonte Orientale}, addressline={largo Donegani 2}, city={Novara}, country={Italy}}
\affiliation[4]{organization={Fondazione Bruno Kessler}, addressline={via Sommarive 18}, city={Povo, Trento}, country={Italy}}
\affiliation[5]{organization={Trento Institute for Fundamental Physics and Applications}, addressline={via Sommarive 14}, city={Povo, Trento}, country={Italy}}
\affiliation[6]{organization={Istituto Nazionale di Fisica Nucleare, Sezione di Perugia}, addressline={via A. Pascoli}, city={Perugia}, country={Italy}}
\affiliation[7]{organization={Consiglio Nazionale delle Ricerche, Istituto Officina dei Materiali}, addressline={via A. Pascoli}, city={Perugia}, country={Italy}}
\affiliation[8]{organization={Universit\`{a} degli Studi di Perugia}, addressline={via G. Duranti 93}, city={Perugia}, country={Italy}}

\fntext[affiliation_lmenzio]{Current affiliation: CERN, Esplanade des Particules 1, Meyrin, Geneva, Switzerland}

\begin{abstract}
The paper reports on the timing resolution achieved with Low-Gain Avalanche Diodes (LGADs), optimised for extreme-fluence conditions, at the DESY Test Beam Facility using 4~GeV/c electrons.
The LGADs adopt an $n$-in-$p$ technology with a $p^{+}$-type boron gain implant, co-implanted with carbon to mitigate acceptor deactivation due to irradiation.
The substrate thickness of the sensors varies from 20 to 45~\micron, with an active area spanning from 0.75~$\times$~0.75 to 1.28~$\times$~1.28~mm$^{2}$.
The experimental setup consisted of a 45~\micron-thick trigger sensor with an active area of 3.6~$\times$~3.6~mm$^{2}$, two device-under-test (DUT) planes, and a Photonis micro-channel plate photomultiplier tube (MCP) as a time reference.
Data taking was performed at the ambient temperature of the facility, at 18$^{\circ}$C.
The gain was measured between 7 and 40 across all non-irradiated sensors in the study.
The timing resolution was calculated from a Gaussian fitting of the difference in times of arrival of a particle at the DUT and the MCP, using the constant fraction discriminator technique.
A timing resolution of 26.4~\ps was achieved in 45~\micron-thick sensors, down to 16.6~\ps in 20~\micron-thick sensors, in the non-irradiation study.
The combination of two 20~\micron-thick LGADs reached a timing resolution of 12.2~\ps.
A set of 30~\micron-thick sensors irradiated with neutrons at fluences between 0.4~$\times$~10$^{15}$ and 2.5~$\times$~10$^{15}$~\mevneut were tested in the beam.
These irradiated sensors achieved a gain between 7 and 30 using a similar apparatus but cooled with solidified CO$_{2}$ to -42$^{\circ}$C.
A timing resolution of 20~\ps was obtained in these irradiated sensors.

\end{abstract}


\end{frontmatter}

\section{Introduction}
\label{introduction}

Future High-Energy Physics (HEP) experiments are expected to reach significantly higher instantaneous luminosities and therefore will require detectors capable of operating under extreme-fluence conditions.
In the CERN High-Luminosity Large Hadron Collider (HL-LHC), the innermost tracking layers of the high-energy detectors, A Toroidal LHC Apparatus (ATLAS) and Compact Muon Solenoid (CMS), will be exposed to a fluence of 3.5~$\times$~10$^{16}$ 1~MeV-neutron-equivalent fluence~(\mevneut)~\cite{Rossi:1471000,LaRosa:2021cof}, while in the Future Circular Collider (FCC-hh), this is expected to reach $\sim$~6~$\times$~10$^{16}$~\mevneut during proton-proton collisions~\cite{ATLAS:2025cya,benedikt2022futurecircularhadroncollider}.

Low-Gain Avalanche Photodiode (LGAD) technology is at the forefront of future detectors, owing to its superior timing resolution~\cite{PELLEGRINI201412} and radiation hardness.
Current LGAD technology can achieve a temporal resolution between 25 and 35~\ps depending on substrate thickness (sub-55~\micron)~\cite{Ferrero_2025}, and can survive in high radiation environments up to $\sim$~2~$\times$~10$^{15}$~\mevneut while maintaining a modest gain~\cite{Padilla:2020sau}.

The LGADs used in the analysis (EXFLU0 and EXFLU1) have been designed to investigate how thin substrates respond to irradiation at extreme fluences, and how timing resolution and gain, among other properties, evolve with substrate thickness and irradiation fluence.
Two batches of sensors manufactured by the Fondazione Bruno Kessler (FBK) were tested.
The EXFLU0 sensors have the same layout and gain implant design as the UFSD3.2 sensor batch~\cite{Tornago_2022} produced at FBK, but on thinner substrates, namely 25 and 35~\micron.
The EXFLU1 sensors have a dedicated sensor layout, with optimised peripheral structures, and further explore the beneficial effects derived from thin substrates down to 15~\micron.
The sensors were irradiated with neutrons in the TRIGA Mark II research nuclear reactor at the Jozef Stefan Institute~\cite{JSI}.
Performance evaluation of EXFLU sensors (EXFLU0 and EXFLU1 batches combined), both before and after irradiation, was completed at the laboratory in Turin and reported in~\cite{Tornago_2022,VS-EXFLU0,ValentinaSola:Frontiers,ALTAMURA2026170799}.

This paper reports on the performance of the EXFLU sensors at the Test Beam Facility at DESY~\cite{DESYTB}.
The EXFLU design is described in detail in Section~\ref{sec:lgad_design}.
The methodology used to test LGAD performance, in the context of gain characterisation and timing analysis, is outlined in Section~\ref{sec:methodology}.
The experimental setup and the approaches to data acquisition, signal analysis, and characterisation of the LGADs, both non-irradiated and irradiated, are explained in Section~\ref{sec:experimental_setup}.
The timing resolution measurements are presented for both non-irradiated and irradiated samples in Section~\ref{sec:results}.
Conclusions on the timing performance of the EXFLU sensors are drawn in Section~\ref{sec:summary}.

\section{The EXFLU LGAD Design}
\label{sec:lgad_design}

\begin{figure}[htbp]
\begin{center}
\includegraphics[width=0.85\textwidth]{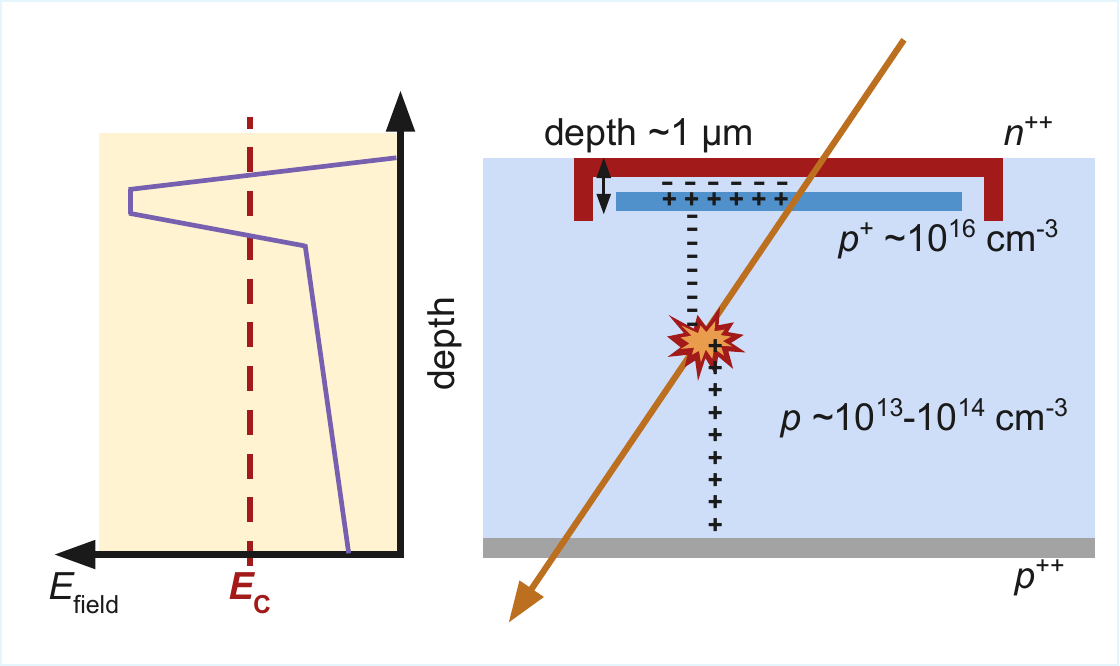}
\end{center}
\caption{Specifications for the LGADs studied: an $n$-in-$p$ design with a peak boron ($p^{+}$-type) doping concentration of $\sim$~10$^{16}$~\atcm in the gain implant.}
\label{fig:lgad_sensor_slice}
\end{figure}

The LGAD design comprises of a silicon pixel sensor with a substrate active thickness of tens of microns (usually $\sim$~50~\micron), implanted with a gain layer $\sim$~1~\micron beneath the surface, operating with a gain of $\sim$~20.
In this study, $n$-in-$p$ LGADs with a $p$-type high-resistivity bulk are used.
The gain implant is boron-doped with a carefully tuned peak $p^{+}$-type concentration of $\sim$~10$^{16}$~\atcm at a depth of $\sim$~1~\micron within the substrate, illustrated in Figure~\ref{fig:lgad_sensor_slice}.
A carbon co-implant is also injected into the gain layer to mitigate the deactivation of boron acceptors due to the effects of irradiation~\cite{WHITE2024169798,Moll:2020kwo}.

Implantation and thermal activation of the carbon and boron acceptors to synthesise the gain layer adhere to two different schemes: low-carbon, low-boron diffusion (CBL), and high-carbon, low-boron diffusion (CHBL)~\cite{Ferrero:2021lwf}.
The CBL mode involves implanting carbon and boron, followed by simultaneous thermal activation at low temperature.
The CHBL mode involves implantation and activation of carbon under a high heat load, followed by implantation and activation of boron at a lower temperature.
%Carbon-boron inactivation effects emerge during thermal activation, and more strongly manifest in CBL than CHBL implantation~\cite{MF:TIPP2021}.

The EXFLU0 batch of sensors was implanted under the CHBL mode.
A single-pad (SP) device measures 1.28~$\times$~1.28~mm$^{2}$ at the surface, while the dual-pad LGAD-PIN (LP) device measures 1.0~$\times$~1.0~mm$^{2}$ on each pad.

\begin{figure}[htbp]
\begin{center}
  \centering
  \begin{subfigure}{0.45\textwidth}
    \includegraphics[width=\linewidth]{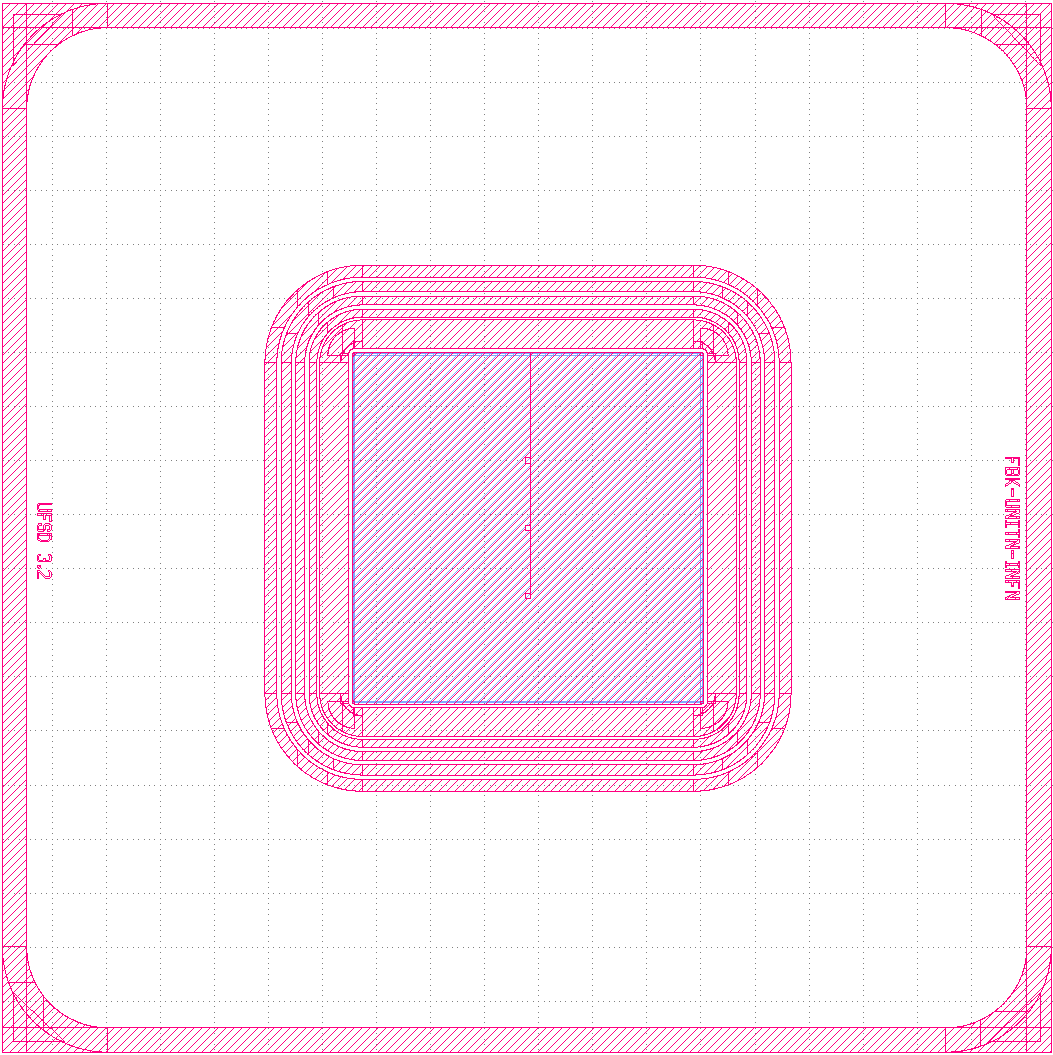}
    \caption{EXFLU0 SP}
  \end{subfigure}
  \hspace*{\fill}
  \begin{subfigure}{0.45\textwidth}
    \includegraphics[width=\linewidth]{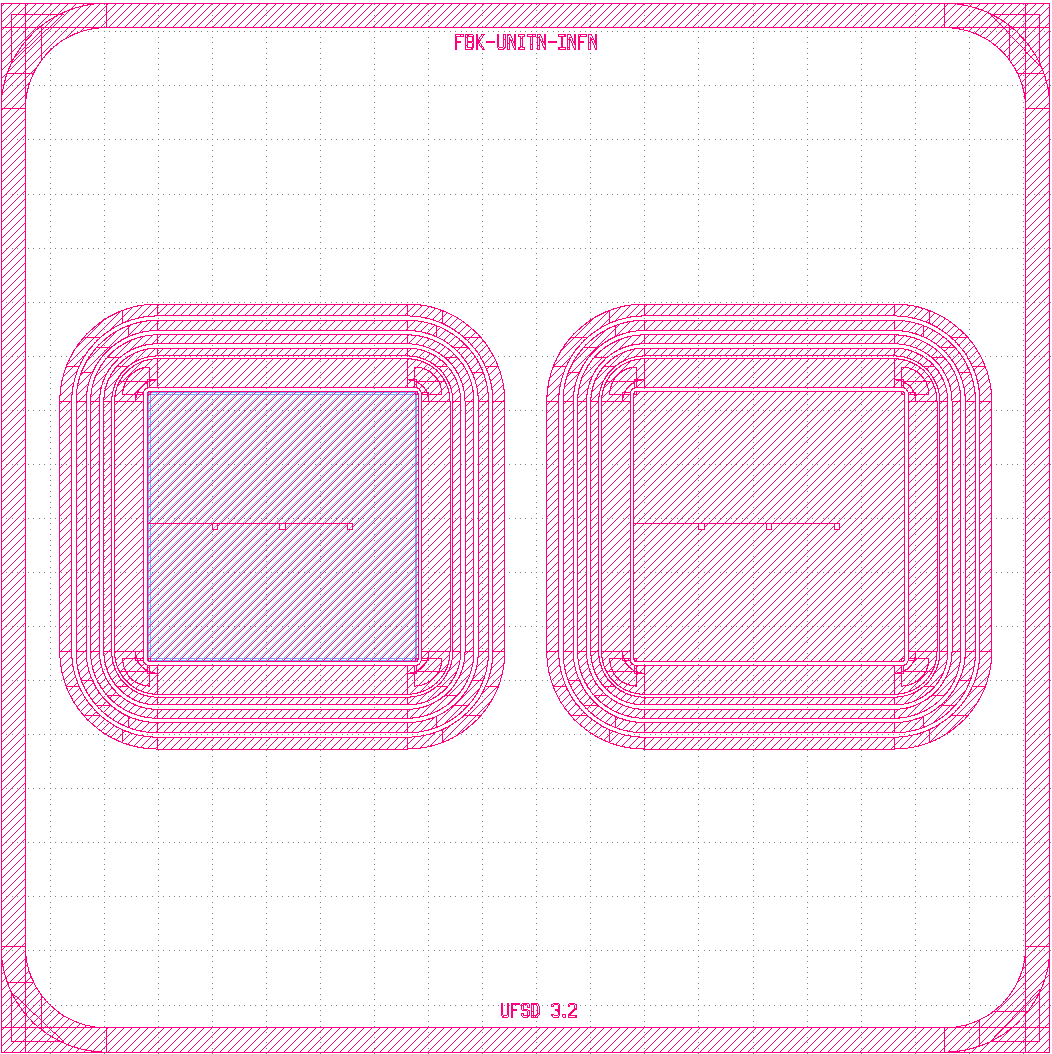}
    \caption{EXFLU0 LP}
  \end{subfigure}
  \newline
  \vspace{0.5cm}
  \newline
  \hspace*{0.085\textwidth}
  \begin{subfigure}{0.27\textwidth}
    \includegraphics[width=\linewidth]{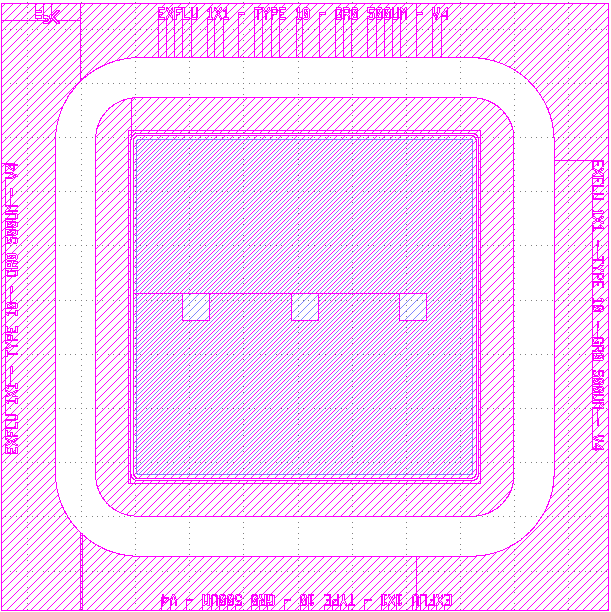}
    \caption{EXFLU1 SP}
  \end{subfigure}
  \hspace*{\fill}
  \begin{subfigure}{0.44\textwidth}
    \includegraphics[width=\linewidth]{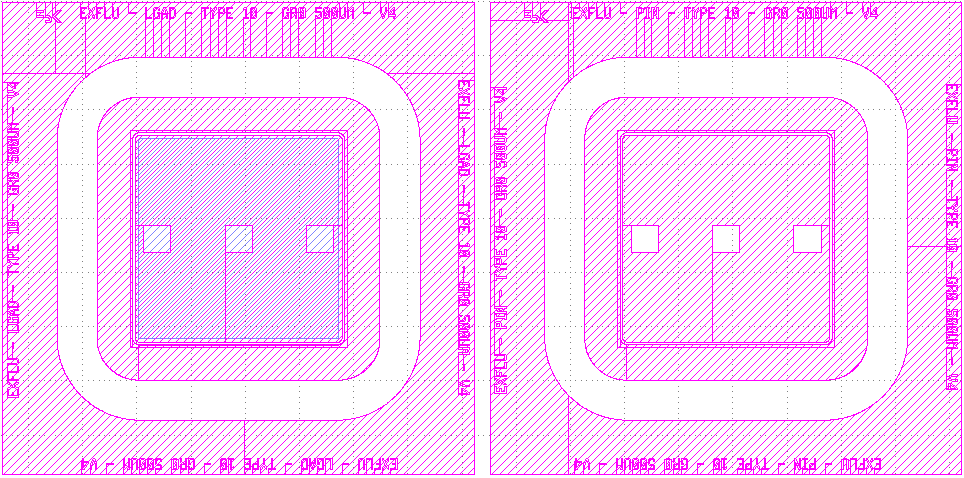}
    \caption{EXFLU1 LP}
  \end{subfigure}
\end{center}

\caption{Schematic images of EXFLU SP and LP devices used for timing performance studies, shown to scale: the EXFLU0 SP (a) and LP (b) devices with an active area of 1.28~$\times$~1.28~mm$^{2}$ and 1.0~$\times$~1.0~mm$^{2}$ per pad, respectively, and the EXFLU1 SP (c) and LP (d) devices with an active area of 1.28~$\times$~1.28~mm$^{2}$ and 0.75~$\times$~0.75~mm$^{2}$ per pad, respectively.}
\label{fig:lgad_pin_single_pad_schematic}
\end{figure}

The EXFLU1 batch of sensors differs from the EXFLU0 batch primarily in implantation, geometry, and periphery design.
The EXFLU1 sensors were implanted under the CBL mode with the exception of one 20~\micron-thick design. 
%The 20~\micron EXFLU1 sensor under CBL implantation has a higher $p^{+}$ dose than the CHBL-activated sensor of the same thickness.
%The measurable consequence of inactivation and the difference in $p^{+}$ dose for the EXFLU1 sensors is the prolongation of sensor bulk depletion under an external reverse bias.
The SP geometry is identical across EXFLU batches, whereas the EXFLU1 LP device has a smaller active area, measuring 0.75~$\times$~0.75~mm$^{2}$ per pad.
The difference in periphery design results in a slightly higher average bulk breakdown voltage for EXFLU1 sensors than for EXFLU0 sensors at a given thickness, but it yields comparable results for the purposes of the timing study.
The EXFLU0 and EXFLU1 schematics for both SP and LP devices are illustrated in Figure~\ref{fig:lgad_pin_single_pad_schematic}.

The nominal thicknesses of the sensors studied at the Test Beam Facility range from 20 to 45~\micron. The effective active substrate thickness is therefore considered to be the nominal thickness minus 2~\micron for the LGADs.
Considering the capacitance of each sensor using a parallel-plate approximation, the LP device with a smaller active area was used for sensors with an active thickness less than 35~\micron.
This ensures that the sensor capacitance is maintained below 5~pF after full sensor depletion.

The boron density in the silicon bulk, \bulkconc, of sensors with thicknesses less than 30~\micron is 1.5~$\times$~10$^{14}$~\bulkconcunit; for 30~\micron-thick sensors, 1.5~$\times$~10$^{13}$~\bulkconcunit; and for thicknesses greater than 30~\micron, $\sim$~1~$\times$~10$^{13}$~\bulkconcunit.

The design specifications, including the $p^{+}$ relative dose (normalised to the FBK reference dose and expressed in arbitrary units) \bulkconc, the devices tested (DUTs), and the pad capacitance are summarised in Table~\ref{tab:tested_lgad_specs}.
In all sensors studied, a carbon dose of 1.0 is co-implanted according to~\cite{Ferrero:2021lwf}.

\newcommand{\groupa}{\cellcolor{thick45colour!60}\textbf{45}}
\newcommand{\groupb}{\cellcolor{thick30colour!60}\textbf{30}}
\newcommand{\groupc}{\cellcolor{thick20chblcolour!60}\textbf{20}}
\newcommand{\groupd}{\cellcolor{thick20colour!60}\textbf{20}}
\newcommand{\groupe}{\cellcolor{thick35colour!60}\textbf{35}}
\newcommand{\groupf}{\cellcolor{thick25colour!60}\textbf{25}}

\begin{table}[t]
\centering
\caption{The design specifications and device types tested, with relative peak gain layer doping concentration, implantation mode, \bulkconc, device type, and sensor capacitance. The EXFLU0 sensors are demarcated as the 35 and 25~\micron substrates, while the EXFLU1 samples correspond to the 45, 30, and 20~\micron substrates. The $p^{+}$ dose is normalised to the FBK reference value. All sensors have a carbon co-implantation of value 1.0, following~\cite{Ferrero:2021lwf}.}
\resizebox{\textwidth}{!}{\begin{tabular}{c c c c c c}
  \hline
  Thick. [\micron] & $p^{+}$ dose [a.u.] & Diffusion & \bulkconc [\bulkconcunit] & DUT & $C_{\rm{DUT}}$ [pF] \\
  \hline
%  \groupe & 0.94 & CHBL & $<$5 $\times$ 10$^{12}$ & SP & 4.8 \\
  \groupe & 0.94 & CHBL & $\sim$ 1 $\times$ 10$^{13}$ & SP & 4.8 \\
  \groupf & 0.94 & CHBL &  1.5 $\times$ 10$^{14}$ & LP & 4.1 \\
  \hline
%  \groupa & 1.14 & CBL  & $<$5 $\times$ 10$^{12}$ & SP & 3.9 \\
  \groupa & 1.16 & CBL  & $\sim$ 1 $\times$ 10$^{13}$ & SP & 3.9 \\
  \groupb & 1.12 & CBL  &  1.5 $\times$ 10$^{13}$ & LP & 1.9 \\
  \groupc & 0.80 & CHBL &  1.5 $\times$ 10$^{14}$ & LP & 2.9 \\
  \groupd & 0.96 & CBL  &  1.5 $\times$ 10$^{14}$ & LP & 2.9 \\
  \hline
\end{tabular}}
\label{tab:tested_lgad_specs}
\end{table}

At least one non-irradiated sensor of each thickness was tested at the Test Beam Facility.
A single CHBL and CBL device were tested in the case of the 20~\micron-thick sensor.
A subset of EXFLU1 30~\micron-thick sensors irradiated to fluences of 0.4~$\times$~10$^{15}$, 0.8~$\times$~10$^{15}$, 1.5~$\times$~10$^{15}$, and 2.5~$\times$~10$^{15}$~\mevneut were also tested.
All measurements of the irradiated sensors were performed after annealing at 60$^{\circ}$C for 80 minutes~\cite{Kramberger:2020ixm}.

The EXFLU1 sensors are well established as having the best acceptor-removal mitigation performance of any thin LGAD produced by FBK to date~\cite{ValentinaSola:Frontiers,ALTAMURA2026170799}.
The irradiated sensors used for this study have demonstrated that the gain performance observed in the non-irradiated sensors can be reproduced up to a fluence of 2.5~$\times$~10$^{15}$~\mevneut when operated at a sufficiently high bias~\cite{WHITE2024169798}.

\section{Methodology}
\label{sec:methodology}

In the context of timing performance, the dominant contributions to the resolution in thin LGADs are the electronic noise, or jitter, which affects the shape of the signal, and the non-uniform energy deposition by ionising particles traversing the sensors~\cite{Ferrero:2021lwf}.

The timing resolution $\sigma_{\textit{t}}$ is therefore given by:
\begin{equation}
    \sigma^{2}_{\textit{t}} = \jitter^{2} + \landau^{2}~,
    \label{eq:timing_resolution_contributions}
\end{equation}
\noindent where \jitter and \landau are the jitter and ionisation contributions, respectively.

The jitter is intrinsically related to the noise in the readout electronics and bandwidth under the relation:
\begin{equation}
    \jitter = \frac{N}{\frac{\textrm{dV}}{\textrm{dt}}} \approx \frac{\risetime}{\frac{S}{N}}~,
    \label{eq:jitter_approximation}
\end{equation}
\noindent where $N$ is the noise, taken as the root-mean-square (RMS) noise of the signal baseline, $\frac{\textrm{dV}}{\textrm{dt}}$ is the slew rate of the oscilloscope, and $S$ and \risetime are the signal magnitude and rise time, respectively.
A smaller electronics bandwidth yields lower noise, whereas a higher slew rate, which minimises jitter, requires a larger bandwidth. Equation~\ref{eq:jitter_approximation} implies that \risetime is directly proportional to \jitter, and therefore, given that faster signals are produced in thinner sensors, $\sigma_{\textit{t}}$ is smaller.

The contribution \landau is invariant of the collected charge but is lower for thinner sensors, given that the faster signals are less sensitive to fluctuations in the signal itself.
The value for \landau could be estimated by rearranging Equation~\ref{eq:timing_resolution_contributions}.
However given that $S$ is proportional to the collected charge $Q$, and that as \risetime and $N$ are independent of $S$, Equation~\ref{eq:jitter_approximation} can be rewritten as:
\begin{equation}
    \begin{aligned}
    \jitter \approx \frac{\risetime}{\frac{S}{N}} \approx \frac{b}{Q}~,~~
    b = \frac{\risetime N}{k}~, 
    \label{eq:jitter_const_derivation}
    \end{aligned}
\end{equation}
\noindent where $k$ is the proportionality constant between $S$ and $Q$, and $b$ is a constant for a given DUT.
Thus, \landau can be extracted by performing a fit of the form of Equation~\ref{eq:timing_resolution_contributions} between the \trdut and $Q$ for each DUT using the formula:
\begin{equation}
  \sigma^{2}_{\textit{t}} = \textit{a}^{2} + \Bigl(\frac{\textit{b}}{\textit{Q}}\Bigr)^{2}~,
\label{eq:landau_term_quad_diff}
\end{equation}
\noindent where $a$ is the constant term corresponding to \landau~\cite{Ferrero2025LGADTiming}.
This fitting approach yields a better estimate of the \landau contribution, given it is fixed for a given device and avoids the propagation of uncertainties in the \jitter approximation when calculating the quadrature difference using Equations~\ref{eq:timing_resolution_contributions} and~\ref{eq:jitter_approximation}.

The charge collected within the sensor is calculated from the fit of a Landau-plus-Gaussian convolution to the distribution of the area of a sensor signal.
The conversion from signal area (in pWb or mV$\cdot$ns) to charge (in fC) is given by:
\begin{equation}
    Q = \frac{\mathcal{A}}{G_{A}R_\textrm{in}}~,
\label{eq:charge_area_conversion}
\end{equation}
\noindent where $\mathcal{A}$ is the baseline-subtracted, time-integrated voltage signal area, and $G_{A}$ and $R_\textrm{in}$ are the amplifier gain factor and input impedance in the readout chain, respectively.
The signal gain is calculated from $Q$ using the formula:
\begin{equation}
  Gain = 100 \frac{Q}{d_{\rm{eff}}}~,
\label{eq:gain_from_charge}
\end{equation}
where $d_{\rm{eff}}$ is the effective active substrate thickness.
This method is equivalent to computing the fraction of $Q$ in the LGAD and in a PIN with identical substrate thickness and area, but without the gain implant, as outlined in~\cite{Meroli_2011}. 

\section{EXFLU operations at the DESY Test Beam Facility}
\label{sec:experimental_setup}

Each sensor sample was read out using a single-channel Santa Cruz (SC) board configured with an $R_\textrm{in}$ of 470~$\Omega$ inverting transimpedance amplifier integrated with an Infineon SiGe RF transistor, capable of supporting bandwidths of $\mathcal{O}$(10~GHz) while providing low noise and high gain~\cite{Ferrero:2021lwf}.
An external amplification stage is incorporated using a commercial low-noise Cividec C1 broadband amplifier with 2~GHz bandwidth and 20~dB gain, corresponding to a $G_{A}$ of 10.
Data acquisition was performed using an eight-channel LeCroy WaveRunner 8208HD oscilloscope, with a 10~GSa/s sampling rate and 2~GHz bandwidth.
The overall system bandwidth, \bw, is $\sim$~1.4~GHz, corresponding to a limit on the measurable inverted signal \risetime = 0.35/\bw of 250~\ps.
The sensors selected for the timing study are not limited by the bandwidth, nor is the minimum measurable \risetime reached for the thinnest sensor substrates, which is expected to be between 300 and 350 ps.

The electron beam momentum at the Test Beam Facility was set to 4~GeV/c, resulting in a particle flux of 1.0~kHz/cm$^{2}$.
The flux is recorded at the facility beam monitor and may underestimate the effective rate.
Due to the different operating biases of the DUTs, two different apparatuses were used: one for non-irradiated and the other for irradiated samples. The details of each setup and LGAD operating requirements are presented in Section~\ref{subsec:nonirrad_setup} for non-irradiated DUTs and Section~\ref{subsec:irrad_setup} for irradiated DUTs, respectively.

\subsection{Non-irradiated setup}
\label{subsec:nonirrad_setup}

The non-irradiated setup was operated at the ambient temperature of the Test Beam Facility, which remained constant at 18$^{\circ}$C during the testing campaign.
The apparatus was set up with a series of three parallel SC boards mounted with sensors for data acquisition, in the following order of beam incidence:
a 45~\micron trigger sensor with active area 3.6~$\times$~3.6~mm$^{2}$ and a minimum inverted signal threshold of 40~mV, operated at 230~V;
the first DUT, DUT~1; 
the second DUT, DUT~2, of the same thickness as DUT~1;
and a Photonis PP2365AC micro-channel plate photomultiplier tube (MCP)~\cite{MCP_photonis} used as a timing reference.
The DUTs were read out simultaneously with the trigger and MCP during data acquisition for each run.
The intrinsic timing resolution of the MCP was calculated at 5~$\pm$~2~\ps using the data taken with 20~\micron-thick sensors, as outlined in Section~\ref{sec:results}.

For each DUT bias point, an acquisition stopping condition of at least 30,000 trigger events was required, equal to about 40~minutes of data taking.
The coincidence rate between the trigger sensor and MCP yielded a 97\% geometric efficiency, and between the DUTs and MCP from 3.4 to 12.7\% coincidence, depending on the sensor geometry.
Across all bias points for a given DUT, data acquisition was performed for up to 8~hours continuously.

\begin{figure}[htbp]
\begin{center}
\includegraphics[width=0.85\textwidth]{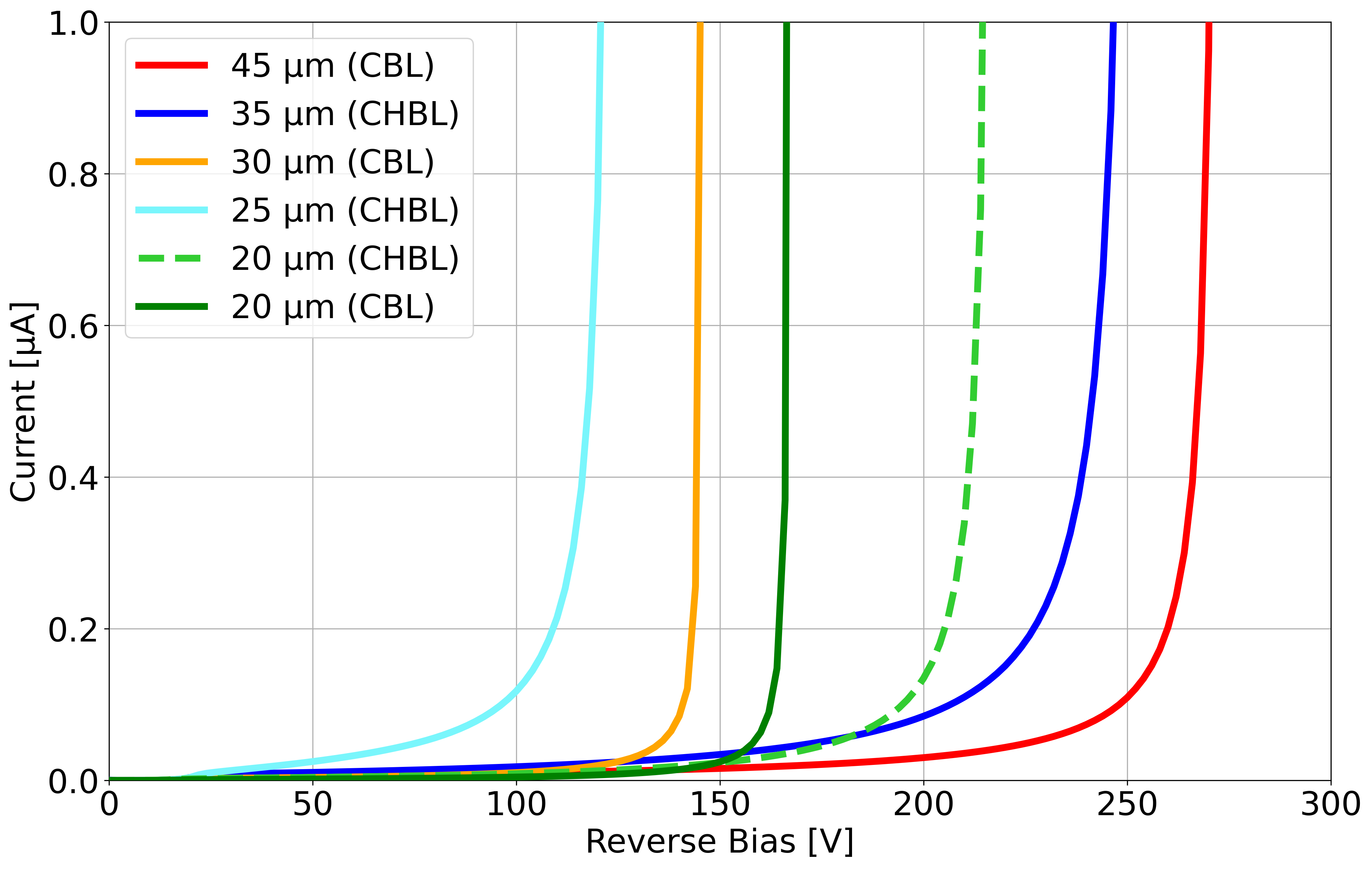}
\end{center}
\caption{The breakdown trends for non-irradiated EXFLU samples measured at 20$^{\circ}$C~\cite{Mulargia2023EXFLU1}.}
\label{fig:IV_EXFLU_summary}
\end{figure}

The reverse bias corresponding to bulk breakdown due to gain in each sample lies within a range of 100~V and 250~V, as shown by the leakage-current-versus-bias curves in Figure~\ref{fig:IV_EXFLU_summary}, measured at 20$^{\circ}$C~\cite{Mulargia2023EXFLU1}.
The range of bias points applied to the DUTs during data taking is selected at varying intervals, starting from a bulk electric field strength of at least 3~\efieldunit up to the DUT breakdown.
The lower bias threshold corresponds to the point at which the electron drift velocity is saturated.

\begin{figure}[htb]
\begin{center}
\includegraphics[width=0.85\textwidth]{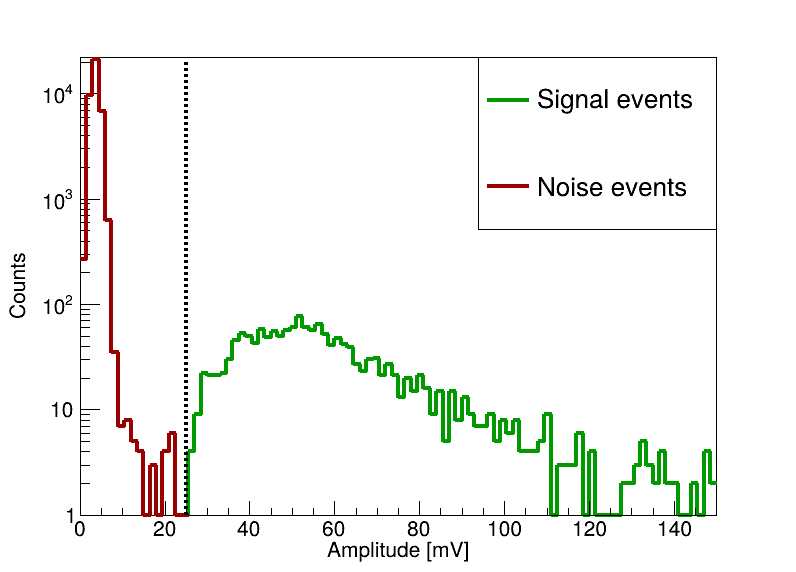}
\end{center}
\caption{The distribution of signal amplitudes for the 20~\micron CBL LGAD operated at 150~V reverse bias and at 18$^{\circ}$C, with the selected signal events indicated above amplitudes of 25~mV. The shape of the signal events, following a Landau distribution convolved with a Gaussian, is typical for all samples.}
\label{fig:signal_v_noise}
\end{figure}

The operating bias ranges ensure a clear separation between the amplitudes of signal and noise events, as demonstrated in Figure~\ref{fig:signal_v_noise} for the CBL-activated 20~\micron device operated under a 150~V bias at 18$^{\circ}$C.
In this example, the region of signal events lies above an amplitude of 25~mV.
The shape of this distribution is typical of the data collected for almost all EXFLU samples: the noise peak width is mostly constant until near the sensor breakdown, while the distribution of signal events broadens, with a higher average amplitude and a more pronounced separation from noise events under higher biases.
The noise in the 45~\micron-thick samples was much larger than in other samples during data acquisition.
Thus, for better comparison with the other thicknesses, the 45~\micron results are replaced with lower-noise data obtained using a 45~\micron-thick sensor tested with a $^{90}$Sr \betanoitalics source of activity 37~kBq at 18$^{\circ}$C.
Most recorded events come from end-point \betanoitalics electrons with an energy of 2.28~MeV, and can be treated as MIP-like.
A single-plane setup is used, with the MCP as the trigger and timing reference, due to the significantly lower energy and flux of ionising particles emitted by the \betanoitalics source.

\begin{figure}[htbp]
\begin{center}
\includegraphics[width=0.85\textwidth]{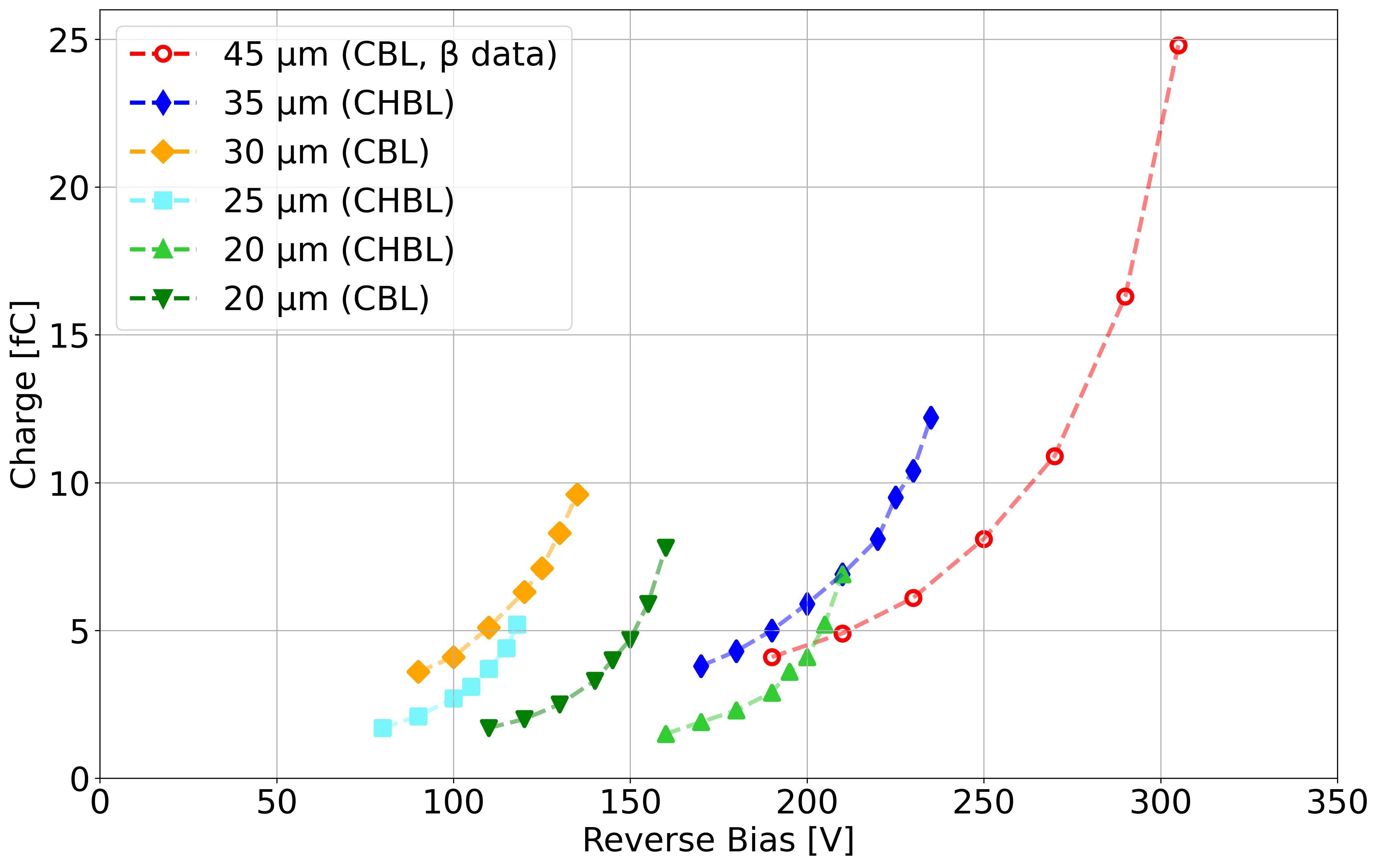}
\end{center}
\caption{Charge collected in non-irradiated EXFLU samples as a function of the reverse bias, recorded at 18$^{\circ}$C.}
\label{fig:charge_v_bias}
%\end{figure}
%\begin{figure}[htbp]
\begin{center}
\includegraphics[width=0.85\textwidth]{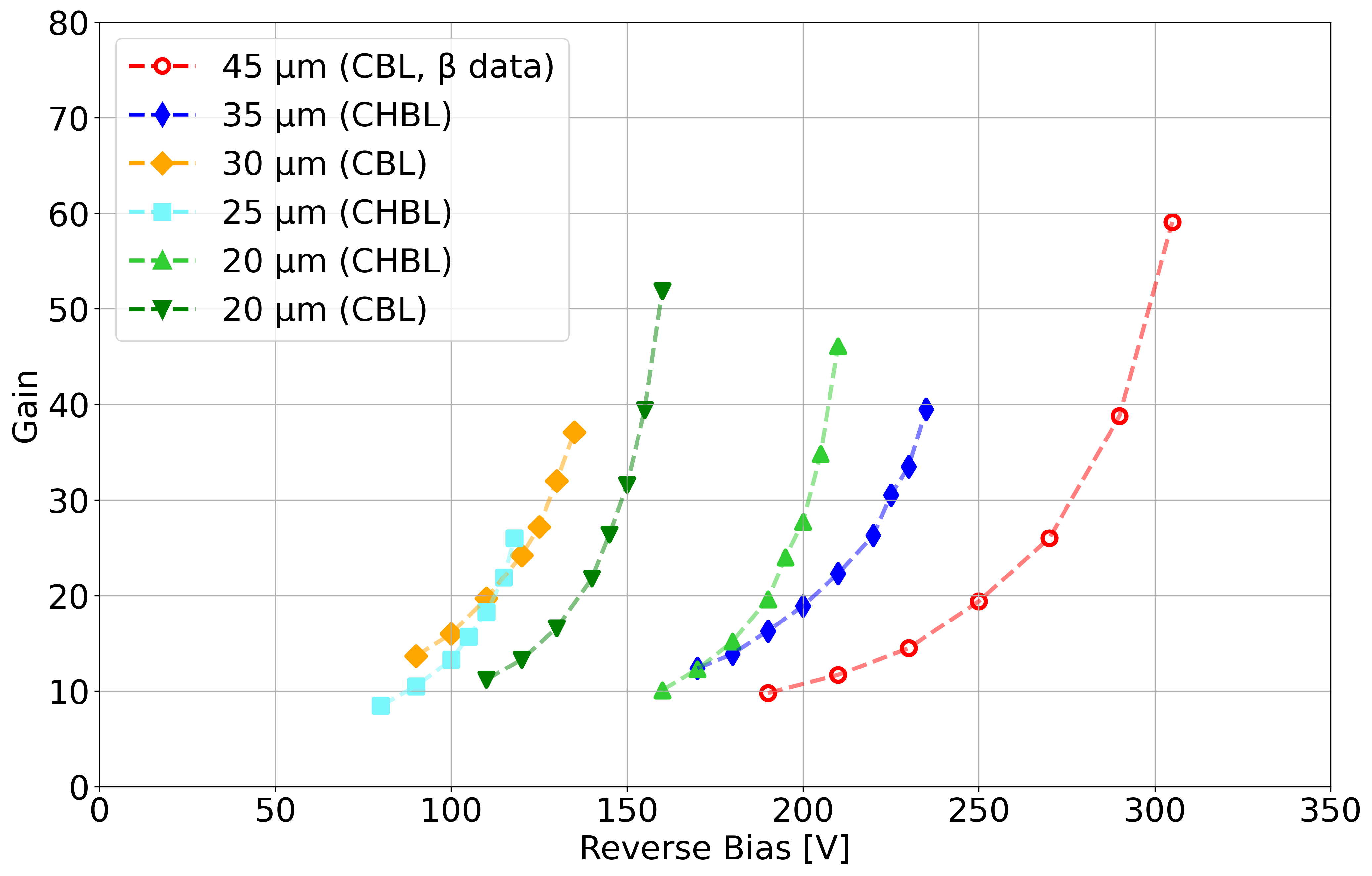}
\end{center}
\caption{Gain measured in non-irradiated EXFLU samples as a function of the reverse bias, recorded at 18$^{\circ}$C, ranging from 7 to 60 largely independent of thickness.}
\label{fig:gain_v_bias}
\end{figure}

The collected charge measured in the non-irradiated samples, calculated following Equation~\ref{eq:charge_area_conversion}, increases with sensor thickness, ranging from 2 to 8~\fC for sensors thinner than 30~\micron, increasing up to 10~\fC for the 30~\micron-thick sensors, and reaching 15~\fC for the 35~\micron-thick sensors, as shown in Figure~\ref{fig:charge_v_bias}.
The gain is observed to vary between 7 and 60 across the EXFLU samples, using Equation~\ref{eq:gain_from_charge}, as presented in Figure~\ref{fig:gain_v_bias}.

\begin{figure}[htb]
\begin{center}
\includegraphics[width=0.85\textwidth]{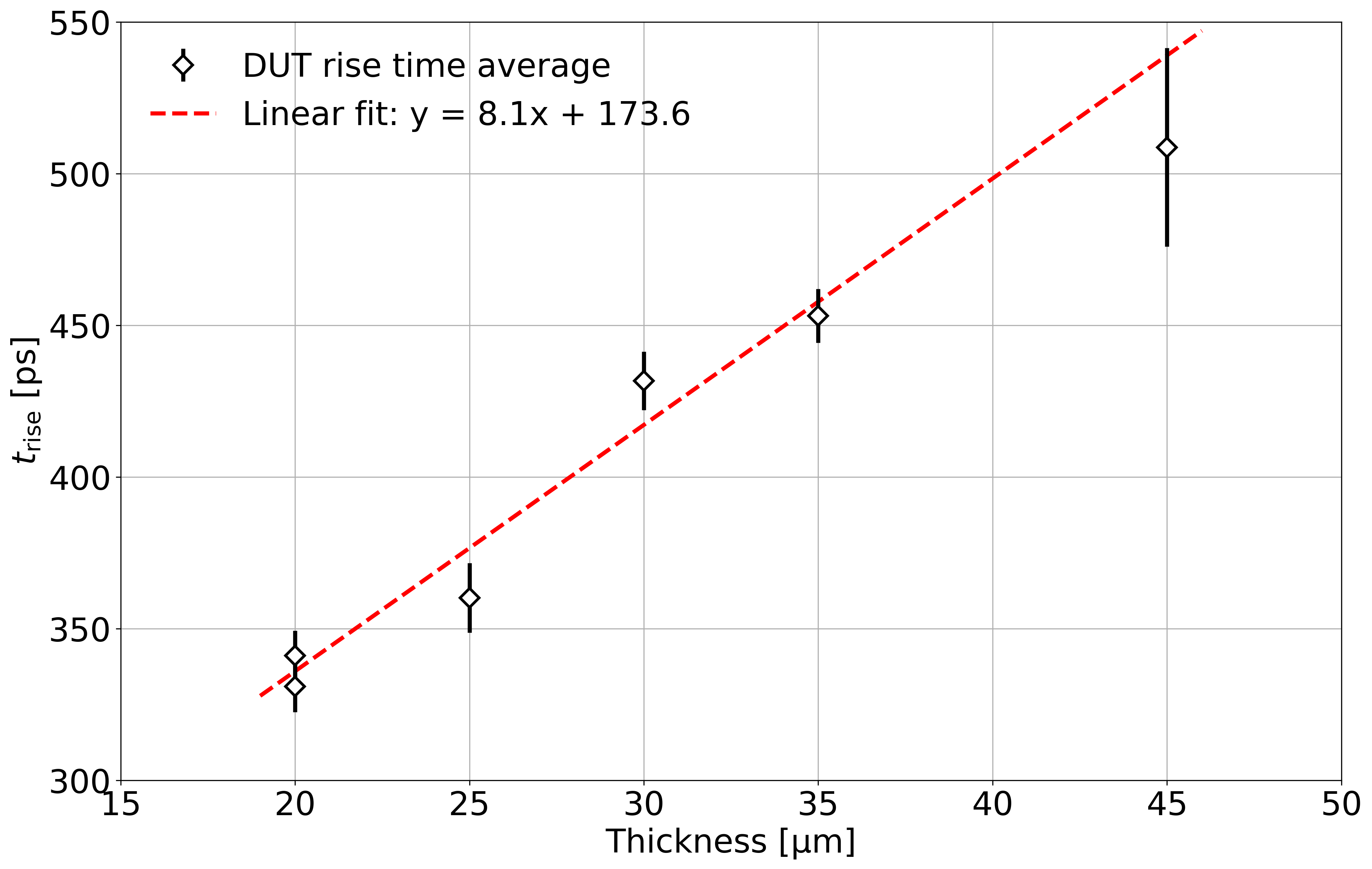}
\end{center}
\caption{The \risetime of each non-irradiated EXFLU sample, taken at the 10\%-90\% interval, demonstrating a linear relationship with the thickness of the sensor.}
\label{fig:rise_time_thickness}
\end{figure}

The \risetime of the non-irradiated LGADs above the saturation threshold for electron drift velocity of 3~V\micron$^{-1}$, taken at the 10\%-90\% interval, is observed to decrease from 600 to 300~\ps linearly with the sensor thickness.
This linear dependence is illustrated in Figure~\ref{fig:rise_time_thickness}, and corresponds to an improving timing resolution in thinner sensors through the \jitter contribution, following Equation~\ref{eq:jitter_approximation}.
The noise is quantified as the signal baseline RMS and is mainly due to the electronics in the setup.
The RMS noise remains stable under increasing reverse bias until sensor breakdown, at which point an increase is observed.
The RMS noise is around 1.8~mV in the 45~\micron-thick sensor, ranges from 1.3 to 1.5~mV in the 25, 30, and 35~\micron-thick sensors, and is between 1.4 and 1.6~mV in 20~\micron-thick samples.

\subsection{Irradiated setup}
\label{subsec:irrad_setup}

The irradiated samples required subzero operating conditions, which was achieved with a polythene cold box filled with solidified CO$_{2}$ to maintain an operating temperature as low as -50$^{\circ}$C.
The temperature was monitored using a DHT11 temperature and humidity sensor integrated with an Arduino system.
The apparatus was set up with a series of five parallel SC boards mounted with sensors for data acquisition, although only three were read out during each period of data acquisition.
This was to minimise the time during which the cold box was opened, thereby preventing ambient heating of the setup.
The same trigger sensor was used in this case, positioned in front of the DUTs but operated at 140~V to account for the difference in temperature.
The same inverted signal threshold of 40~mV was used.

A temperature variation of -50 to -42$^{\circ}$C was recorded by the DHT11 Arduino sensor, with the exception of the sample irradiated at 1.5~$\times$~10$^{15}$~\mevneut, which was measured from -50 to -35$^{\circ}$C.
The variation in temperature was due to the reduction in CO$_{2}$ volume caused by sublimation during the prolonged data-taking periods.

\begin{figure}[htbp]
\begin{center}
\includegraphics[width=0.95\textwidth]{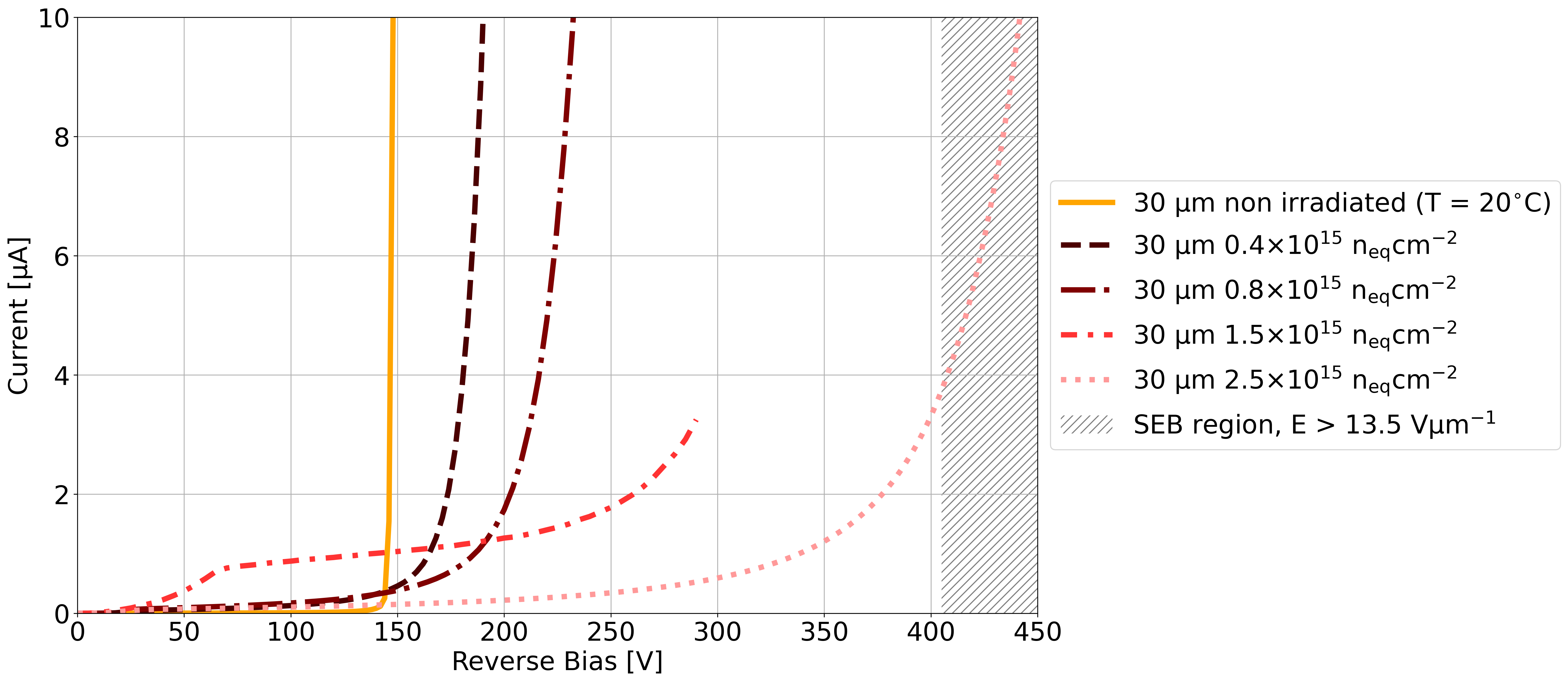}
\end{center}
\caption{The breakdown curves for the irradiated 30~\micron-thick samples, performed at -20$^{\circ}$C, highlighting the increasing breakdown with higher fluences due to the onset of acceptor removal~\cite{WHITE2024169798,Mulargia2023EXFLU1}. The SEB region is also shown here, above the 405~V threshold for 30~\micron-thick sensors~\cite{Ferrero:2025xby}.}
\label{fig:IV_irrad_summary}
\end{figure}

The breakdown trends for irradiated sensors are obtained from measuring the leakage current as a function of the reverse bias at -20$^{\circ}$C.
Operating sensors at low temperatures is necessary to mitigate the increase in leakage current induced by irradiation and to prevent early breakdown.
The irradiated 30~\micron-thick sensors operate at higher biases than the non-irradiated sample, between 200 and 400~V.
The breakdown point of each sensor increases with the level of irradiation, highlighted in Figure~\ref{fig:IV_irrad_summary}.
The single-event burnout (SEB) limit for a 30~\micron-thick substrate is 405~V, considering a maximum electric-field strength of 13.5~\efieldunit~\cite{Ferrero:2025xby}.
The acceptor removal coefficient in 30~\micron EXFLU1 sensors was estimated at 1.22~$\pm$~0.10~$\times$~10$^{16}$~\invmevneut~\cite{ALTAMURA2026170799}, and is the most radiation-hard gain implant design produced by FBK to date.

The collected charge measured in the irradiated 30~\micron-thick sensors under a given bias is sensitive to temperature changes during data taking.
Therefore the charge, and hence the gain as defined in Equation~\ref{eq:gain_from_charge}, is shifted by incrementing the effective reverse bias by 2~V per 1$^{\circ}$C to obtain a bias range at a fixed temperature of -42$^{\circ}$C~\cite{Ferrero:2021lwf}, corresponding to the highest common temperature of all the irradiated sensors.
The temperature correction is small and does not affect the physical quantities of the sensor.

\begin{figure}[htbp]
\begin{center}
\includegraphics[width=0.85\textwidth]{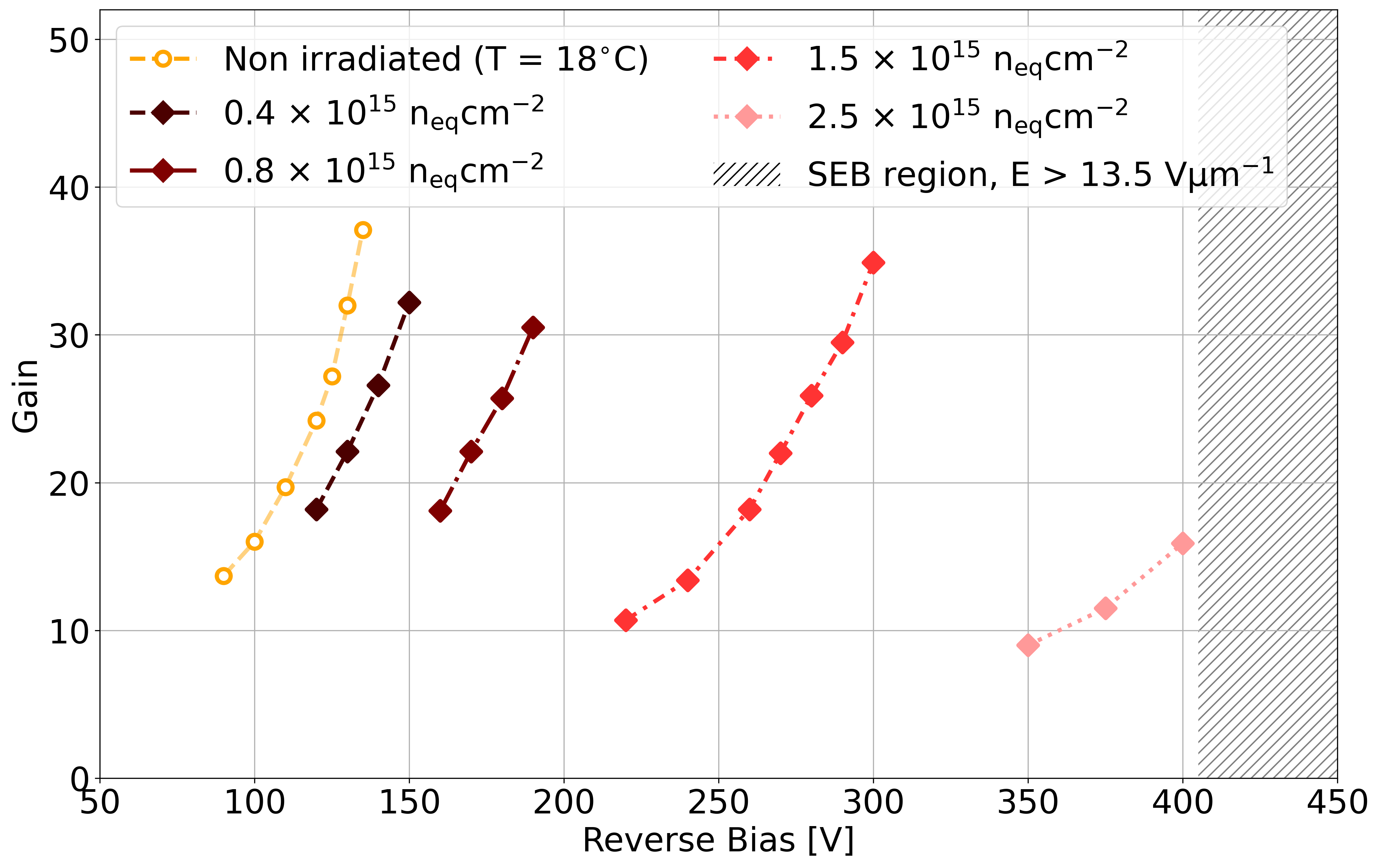}
\end{center}
\caption{Gain measured in the 30~\micron EXFLU samples as a function of the reverse bias, recorded between -50 and -35$^{\circ}$C in the irradiated samples. The non-irradiated sample, tested at 18$^{\circ}$C, is shown for reference. The signal multiplication ranges from 10 to 35.}
\label{fig:gain_v_bias_irrad}
%\end{figure}
%\begin{figure}[htbp]
\begin{center}
\includegraphics[width=0.85\textwidth]{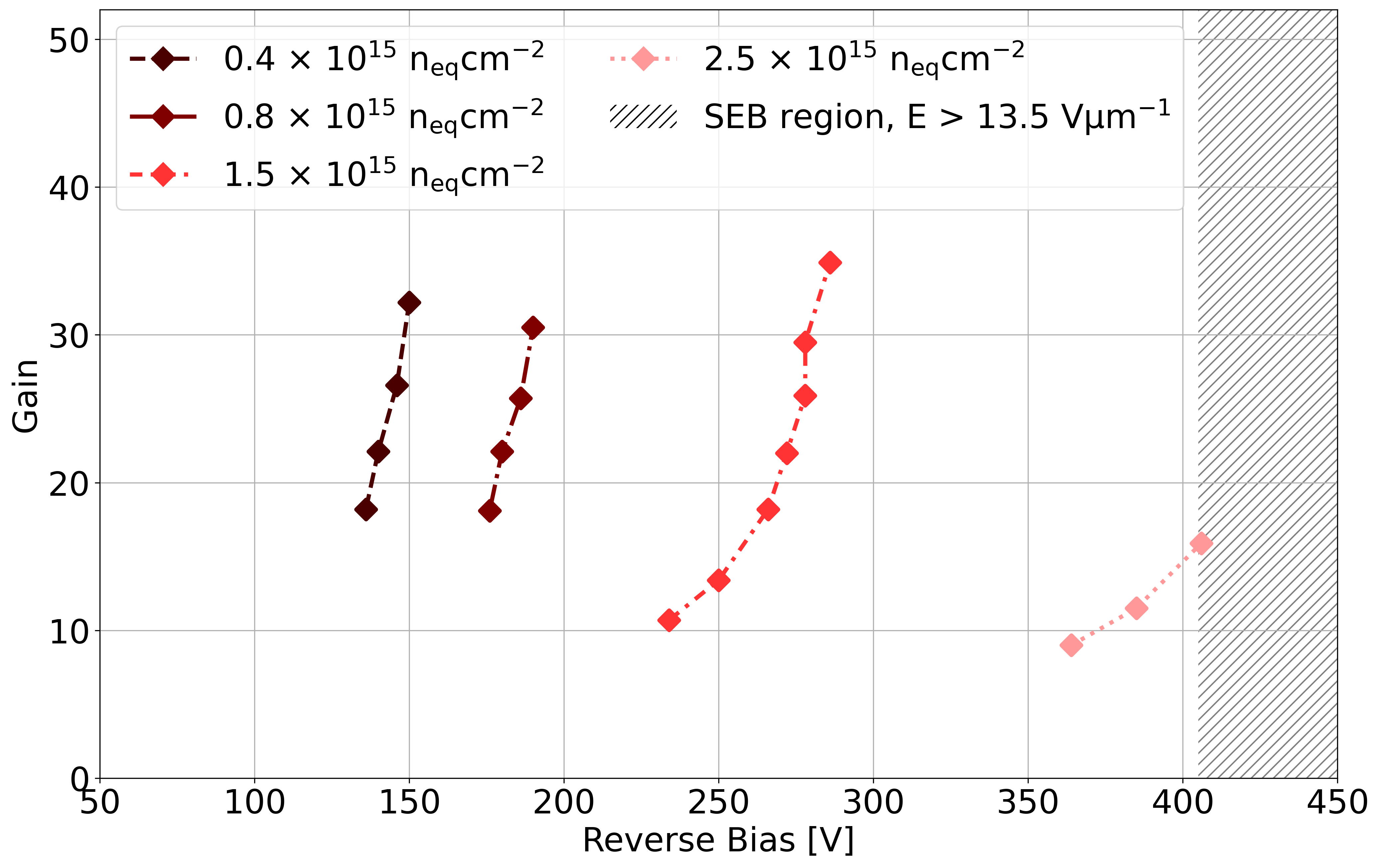}
\end{center}
\caption{Gain measured in the irradiated 30~\micron EXFLU samples as a function of the reverse bias, corrected to -42$^{\circ}$C. The signal multiplication ranges from 10 to 35.}
\label{fig:gain_v_bias_irrad_tempcorr}
\end{figure}

The collected charge varies from 2 to 9~\fC in sensors irradiated up to 1.5~$\times$~10$^{15}$~\mevneut, and between 2 and 4.5~\fC in the sensor irradiated at 2.5~$\times$~10$^{15}$~\mevneut at biases below the SEB limit.
These correspond to a gain of between 10 and 35, and between 10 and 15, respectively, as demonstrated in Figure~\ref{fig:gain_v_bias_irrad} for charge values at the temperature at which the data was recorded, and in Figure~\ref{fig:gain_v_bias_irrad_tempcorr} for data corrected to -42$^{\circ}$C.

The \risetime decreases marginally with fluence in the irradiated 30~\micron-thick samples, due to the increased operating bias.
The RMS noise in these samples ranges from 1.2 to 1.4~mV, except for the sample irradiated at 1.5~$\times$~10$^{15}$~\mevneut, which reached 2.2~mV near breakdown.

\section{Timing performance from 4~GeV/c electron-beam tests}
\label{sec:results}

The timing resolution is calculated using a constant fraction discriminator (CFD) algorithm, which sets the time of arrival of an electron at the DUT and the MCP as the time for which the signal is at least 30\% of the amplitude recorded at each stage.
In addition to the DUT signal event selection described by Figure~\ref{fig:signal_v_noise}, a selection is applied to MCP events.
This requires MCP events with amplitudes above the threshold of 40~mV but below the oscilloscope vertical scale saturation, set to 540~mV.
Only coincident events between the DUT and MCP are used for this calculation following the selections, while the sensitivity of the timing resolution to additional event selections was found to be negligible.
The optimal CFD thresholds were found to correspond to the time of arrival at the 30\% level for all tested sensors.
The distribution of the difference in CFD values from the DUT and MCP events is then fitted with a Gaussian function.
The standard deviation of this fit is the timing resolution of the combined DUT-MCP system, \trdutmcp.
The timing resolution of the DUT, \trdut, is calculated from the quadrature difference:
\begin{equation}
  \trdut = \sqrt{\trdutmcpsquared - \trmcpsquared}~,
\label{eq:time_res_quad_diff}
\end{equation}
\noindent where \trmcp is the contribution due to the intrinsic timing resolution of the MCP.

The MCP timing resolution is computed using the data collected with the EXFLU 20~\micron-thick sensors, which give a very similar timing performance to one another.
This assumes a tri-plane setup whereby the CFD value at 30\% is measured for each pairing of planes, which gives the difference in time of arrival between each DUT and the MCP, as well as the difference between the two DUTs.
From these differences, \trdutmcpsquaredone, \trdutmcpsquaredtwo, and \trdutdutsquared are extracted with a Gaussian fit.
The value for \trmcp is computed in this case using the equation:
\begin{equation}
  \trmcp = \frac{1}{\sqrt{2}}\sqrt{\trdutmcpsquaredone + \trdutmcpsquaredtwo - \trdutdutsquared}~,
\label{eq:triplane_trdut}
\end{equation}
\noindent where DUT~1 is the 20~\micron CHBL sensor of higher operating bias, DUT~2 is the 20~\micron CBL sensor of lower operating bias, and the terms above exploit the quadrature relations:
\begin{equation}
  \begin{aligned}
  \trdutmcpsquaredone &= \trdutsquaredone + \trmcpsquared, \\
  \trdutmcpsquaredtwo &= \trdutsquredtwo + \trmcpsquared, \\
  \trdutdutsquared &= \trdutsquaredone + \trdutsquredtwo.
  \label{eq:triplane_quadrature_relations}
  \end{aligned}
\end{equation}

The \trmcp value is estimated from multiple sets of data, since data acquisition was performed for several bias points, and an average timing resolution of 5~$\pm$~2~\ps is achieved.
The uncertainty of 2~\ps is obtained from the standard deviation in the \trmcp values across all bias points.

\subsection{Non-irradiated results}
\label{subsec:nonirrad_results}

\begin{figure}[htb]
\begin{center}
\includegraphics[width=0.85\textwidth]{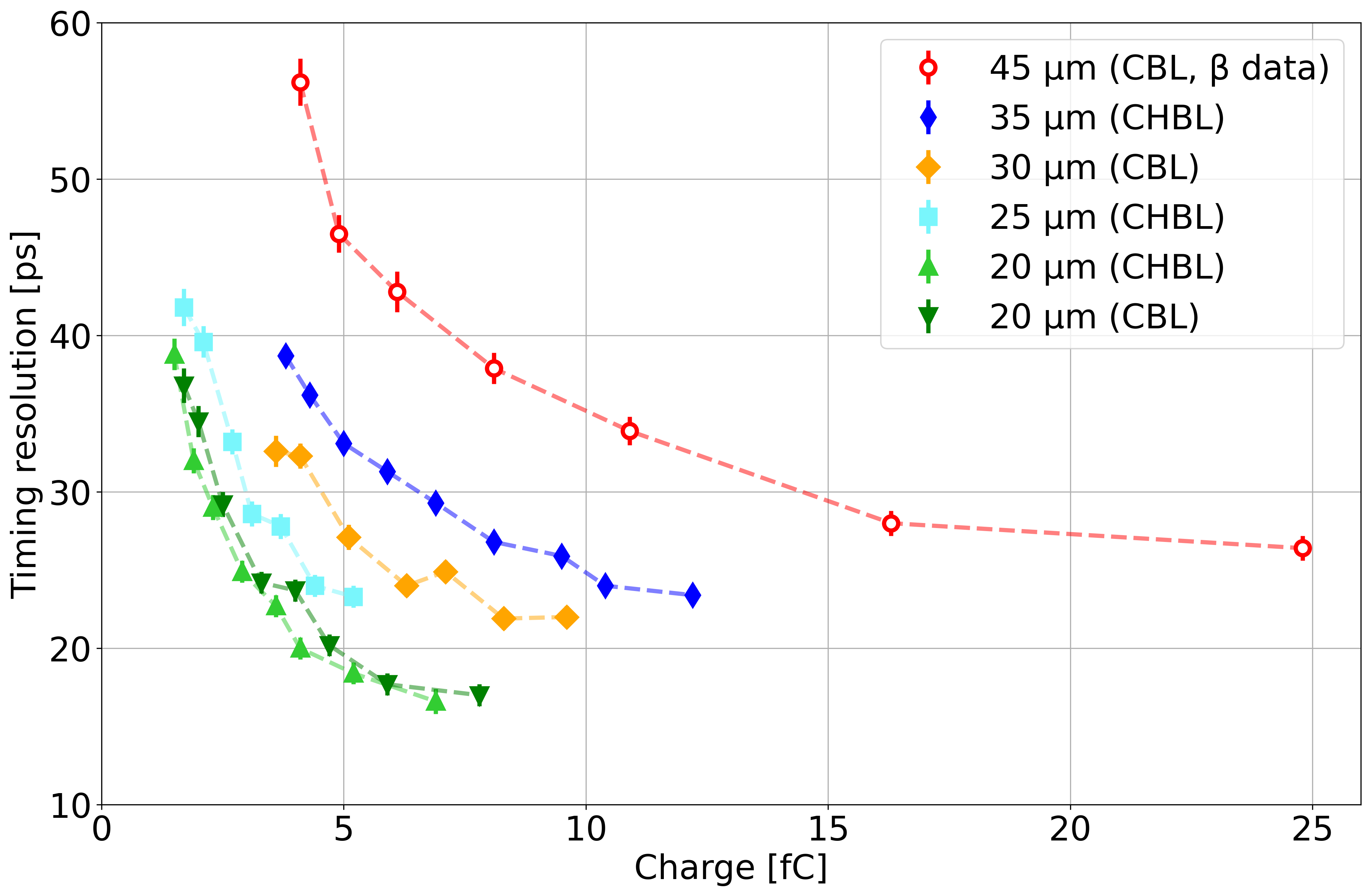}
\end{center}
\caption{The \trdut values across all non-irradiated EXFLU samples as a function of the charge collected, illustrating the effect of sensor thickness on timing precision. The 45~\micron result is recorded using \betanoitalics-source electrons.}
\label{fig:nonirrad_timeres}
\end{figure}

The timing resolution \trdut of all non-irradiated EXFLU samples studied at the Test Beam Facility is presented in Figure~\ref{fig:nonirrad_timeres} as a function of the collected charge, with the lowest measured value for timing resolution of each sensor thickness, \trdutmin, given in Table~\ref{tab:lowest_measured_timeres}.
The values for \trdut use the single-plane method and Equation~\ref{eq:time_res_quad_diff}, for a \trmcp of 5~$\pm$~2~\ps.
The 45~\micron-thick sample is measured using the \betanoitalics setup, which although performed at the same temperature had a lower signal-event yield compared to data taken at the Beam Test Facility.
The MCP timing reference in this case reported a \trmcp of 10~$\pm$~4~ps due to the lower incident electron energies, extracted using the same method as performed for the test beam data.
The most evident trend is the effect of sensor thickness on \trdut, corresponding to the linear decrease in \risetime with substrate thickness highlighted in Figure~\ref{fig:rise_time_thickness}.

\begin{table}[htbp]
\centering
\caption{The lowest measured \trdut value achieved for each non-irradiated EXFLU sensor thickness.}
\begin{tabular}{cc}
Thickness [\micron] & \trdutmin [ps] \\
\hline
45 & 26.4 $\pm$ 1.7 \\
35 & 23.4 $\pm$ 0.5 \\
30 & 22.0 $\pm$ 0.7 \\
25 & 23.3 $\pm$ 0.8 \\
20 & 16.6 $\pm$ 0.7 \\
\end{tabular}
\label{tab:lowest_measured_timeres}
\end{table}

The statistical uncertainty on \trdut calculated directly from the Gaussian fitting is $\lesssim$2\%.
The number of selected signal events in the calculation of \trdut, both for LP and SP devices, is sufficient that the statistical contribution is negligible in the overall uncertainty.
Another uncertainty contribution comes from the \trmcp measurement, although the 2~\ps uncertainty is as reasonable an estimate as possible with the available samples, given the low \trmcp of 5~\ps.
The uncertainties on the \trdut estimates are extracted from the uncertainty on the width of the Gaussian fit to the difference in the CFD of the DUT and MCP, after which the 2~\ps uncertainty in the \trmcp is propagated.
These uncertainties are minimal, varying across all bias points and thicknesses between 0.5 and 1.1~\ps for the non-irradiated samples.
The 45~\micron-thick sample tested at the \betanoitalics setup has an uncertainty range of 0.8 to 1.5~\ps.

A test was performed to verify the reliability in determining the signal amplitude due to the sampling rate of the oscilloscope, which directly affects the CFD.
A large subset of signal data with amplitude within a 1~mV range was used.
The functional form used to extract the amplitude was compared across different distributions, such as symmetric Gaussian, Lorentzian, parabolic, and Voigtian fits, and asymmetric Landau and skewed-Gaussian fits.
The effect was found to be negligible on the overall timing resolution.
Similarly, the means of extracting the CFD value were compared between interpolating a data point via a linear fit to the two closest sampling points at the 30\% threshold recorded by the oscilloscope, and using a spline fit along the rising edge of the signal shape.
Again, there was no discernible effect on the timing resolution.

\begin{figure}[htb]
\begin{center}
\includegraphics[width=0.85\textwidth]{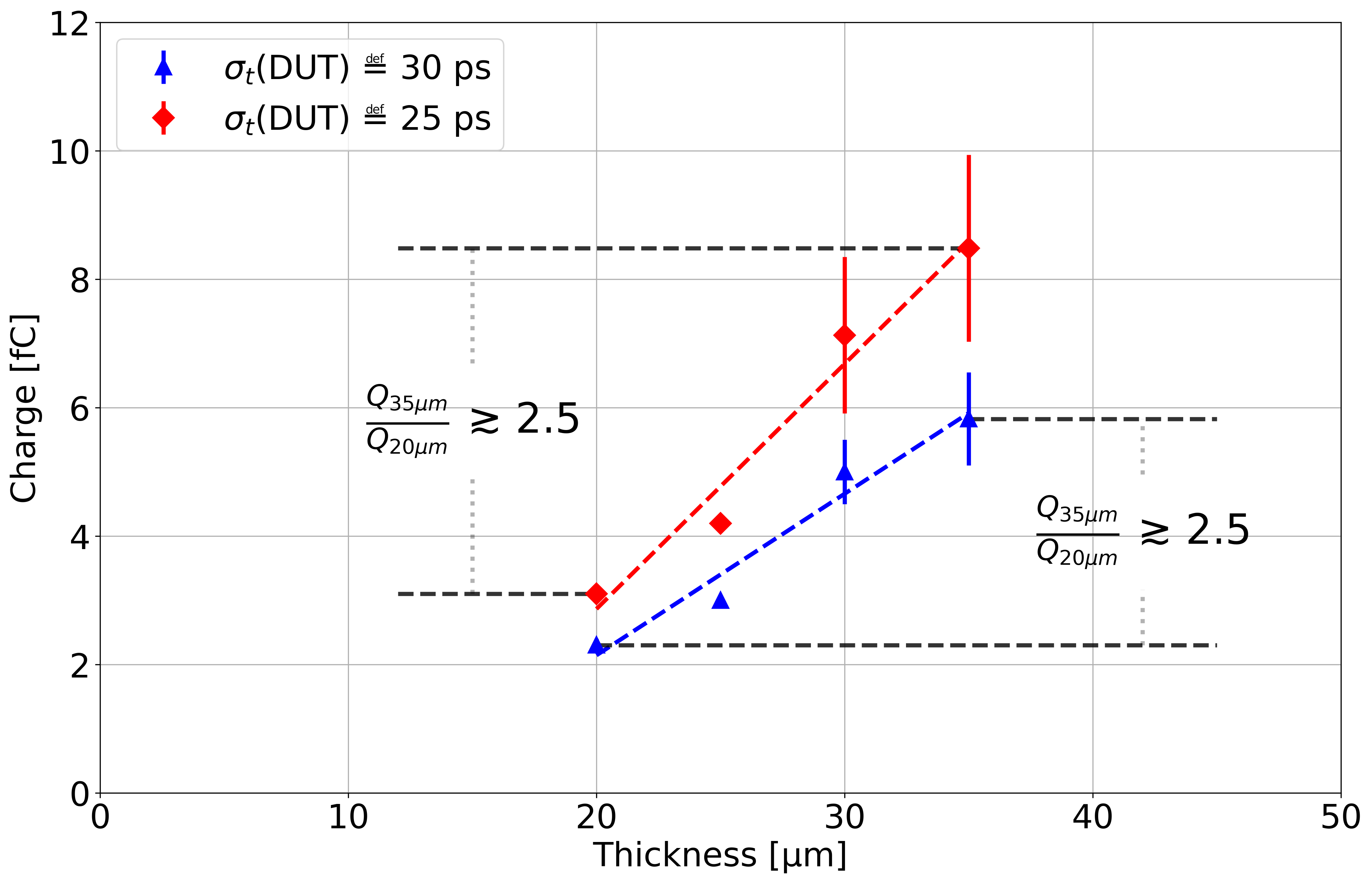}
\end{center}
\caption{The minimum collected charge required to achieve a fixed \trdut value as a function of sensor thickness, highlighting the sensitivity between different EXFLU samples for a signal measured with a given precision.}
\label{fig:charge_thickness}
\end{figure}

The amount of collected charge necessary to reach a given \trdut is also observed to decrease linearly with the sensor thickness, illustrated in Figure~\ref{fig:charge_thickness}.
In this case, the CBL-activated 20~\micron device is used as the reference, although the performance of the two 20~\micron-thick sensors is nearly identical.
The data demonstrate that, for a sub-30~\ps \trdut, the minimum collected charge is more than a factor of 2.5 smaller between the 35~\micron and 20~\micron-thick sensors, decreasing from $\sim$~6~\fC to $\sim$~2~\fC.
The corresponding minimum charge for a 45~\micron-thick sensor is $\sim$~15~\fC.

\begin{figure}[htb]
\begin{center}
\includegraphics[width=0.85\textwidth]{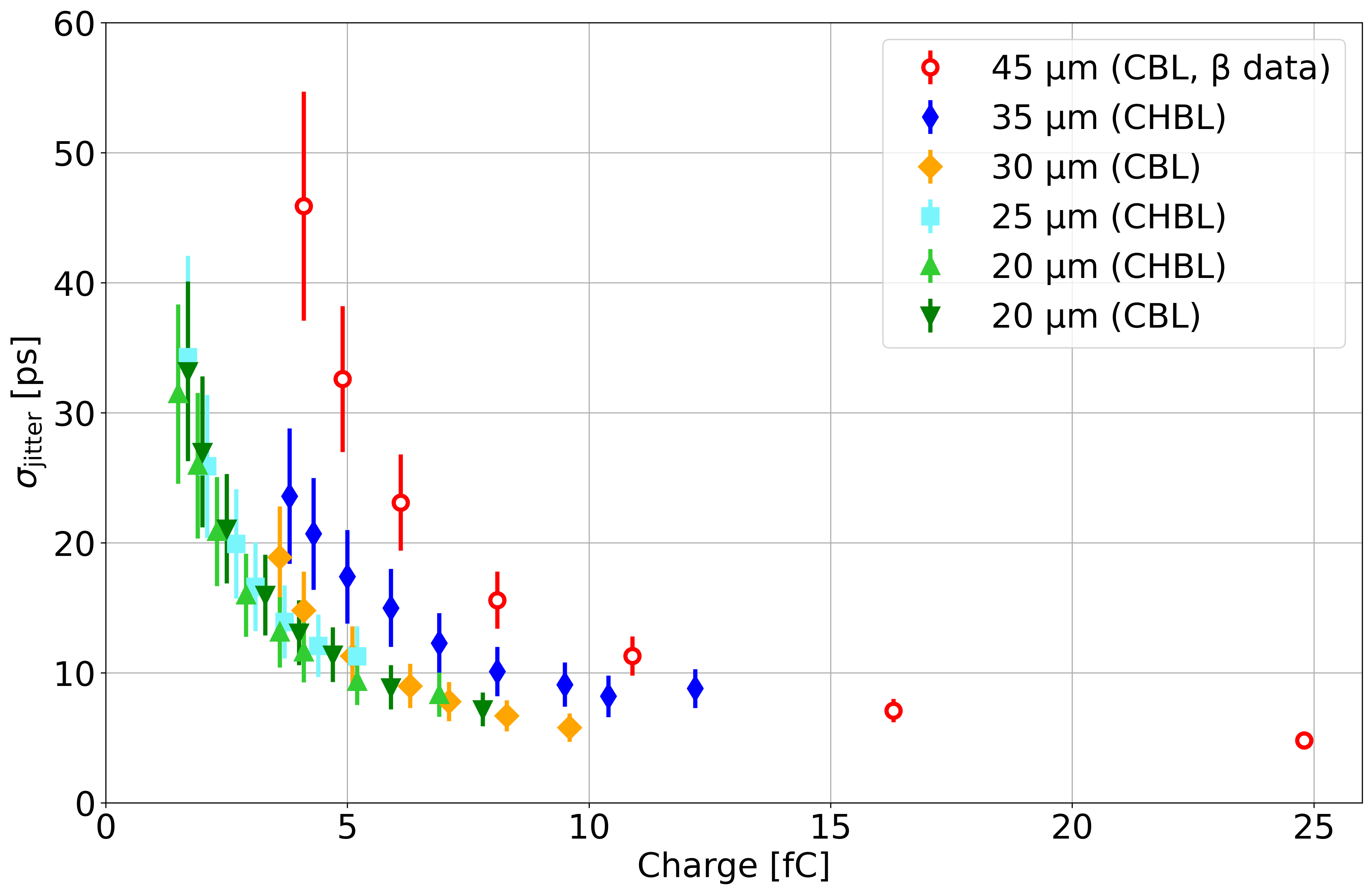}
\end{center}
\caption{The \jitter contribution as a function of the collected charge by EXFLU samples. A sub-10~\ps \jitter is achieved in all sensors of thickness below 45~\micron.}
\label{fig:jitter_charge}
\end{figure}

The timing resolution measurements are composed of \jitter and \landau contributions, described by Equation~\ref{eq:timing_resolution_contributions}~\cite{Ferrero:2021lwf}.
The \jitter contribution, measured using the approximation in Equation~\ref{eq:jitter_approximation}, is $\lesssim$10~\ps in sensors below 45~\micron and $\lesssim$20~\ps in the 45~\micron-thick sample, as shown in Figure~\ref{fig:jitter_charge}.

\begin{figure}[htb]
\begin{center}
\includegraphics[width=0.85\textwidth]{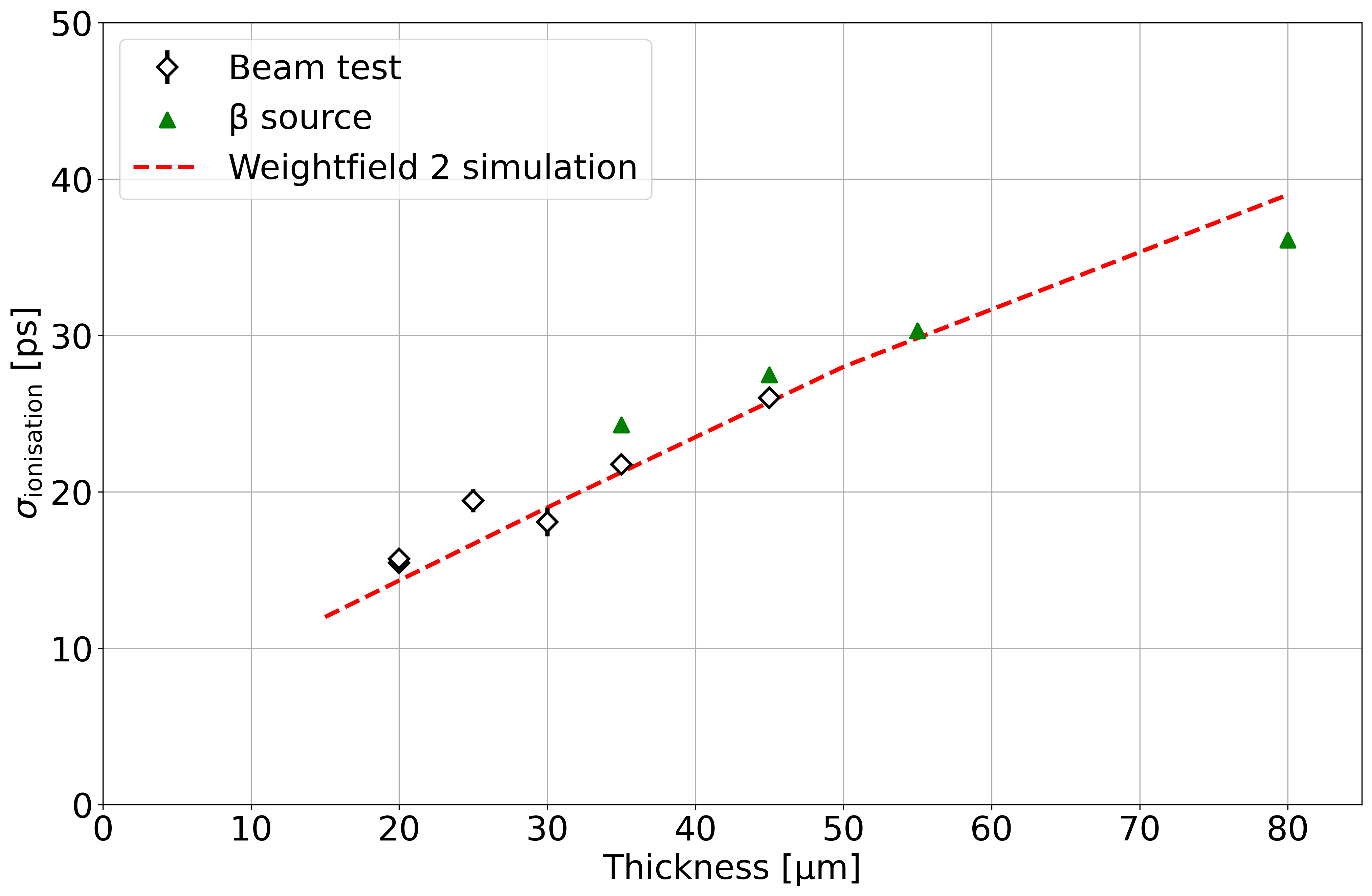}
\end{center}
\caption{The \landau contribution as a function of thickness using test-beam data and from samples measured with a \betanoitalics source. The observations are in good agreement with the simulated \landau behaviour using Weightfield2~\cite{Cenna:2015xtw}.}
\label{fig:landau_thickness}
\end{figure}

The \landau term is measured with the fit described in Equation~\ref{eq:landau_term_quad_diff}, and is observed to decrease from 25~\ps for the 45~\micron-thick sample to 15~\ps in the 20~\micron-thick samples, shown in Figure~\ref{fig:landau_thickness}.
A consistency check was performed by inputting the \trdut and \jitter values at the bias point of minimum \jitter for each DUT thickness into Equation~\ref{eq:timing_resolution_contributions} to extract \landau.
The \landau results obtained in this case were near-identical to those obtained using Equation~\ref{eq:landau_term_quad_diff}, but with larger uncertainties due to the larger uncertainty in the \jitter terms.
The \landau results are corroborated by previous measurements of sensors with thicknesses ranging from 35~\micron to 80~\micron using the \betanoitalics setup~\cite{Siviero:2021olm}, and are in good agreement with the predicted \landau trend from Weightfield2 simulations~\cite{Cenna:2015xtw}.
This is in agreement with the expectation that thinner sensors, which produce signals with steeper rising edges, are less affected by fluctuations due to non-uniform ionisation of traversing particles within the sensor.

\begin{figure}[htb]
\begin{center}
\includegraphics[width=0.85\textwidth]{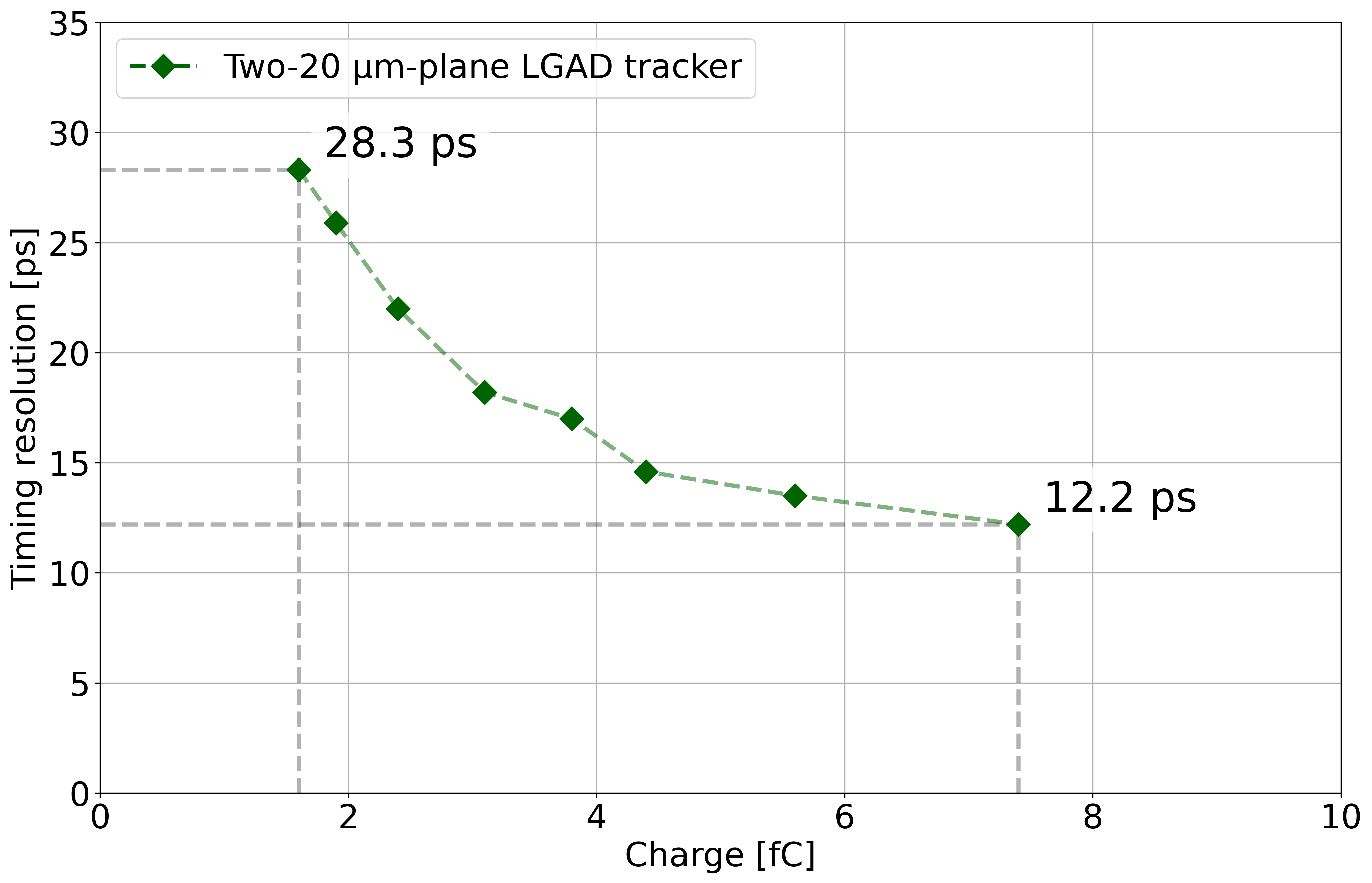}
\end{center}
\caption{The timing resolution of a two-plane LGAD tracker setup with the 20~\micron-thick samples, where the average time of arrival of each DUT is used to compute the value for \trdutdut. The lowest \trdutdut achieved by the two-plane tracking system is 12.2~\ps.}
\label{fig:biplane_timeres}
\end{figure}

The timing resolution of a two-plane LGAD tracker system using the 20~\micron DUTs has been evaluated to verify the overall $\sigma_t$ scaling by a factor of 1/$\sqrt{N}$, where $N$ is the number of tracking planes.
The time of the track using the CFD at the 30\% level for the two 20~\micron DUTs, \tcfddutone and \tcfdduttwo, is given by:
\begin{equation}
  [\tcfddutone + \tcfdduttwo] / 2~.
\label{eq:lgad_tracker}
\end{equation}
A Gaussian fit to the time of the track yields a \trdutdut of 12.2~\ps, once the \trmcp contribution is subtracted.
The results of the LGAD tracker for the 20~\micron-thick samples are shown against collected charge in Figure~\ref{fig:biplane_timeres}.

\subsection{Irradiated results}
\label{subsec:irrad_results}

\begin{figure}[htbp]
\begin{center}
\includegraphics[width=0.83\textwidth]{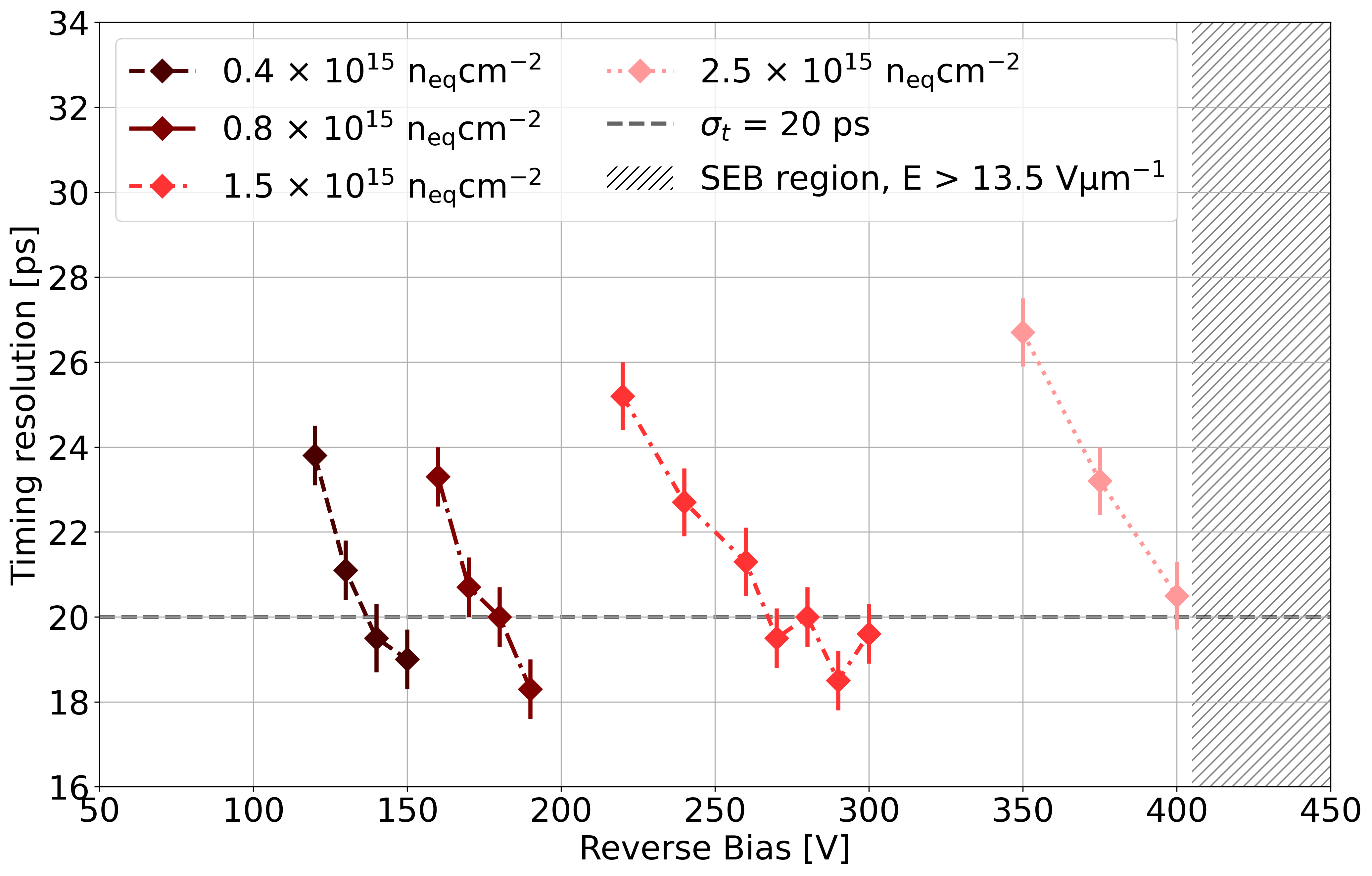}
\end{center}
\caption{The timing resolution of the irradiated 30~\micron-thick sensors as a function of reverse bias, using data obtained between -50 and -35$^{\circ}$C, with a 20~\ps baseline shown for reference.}
\label{fig:irrad_timeres}
%\end{figure}
%\begin{figure}[htbp]
\begin{center}
\includegraphics[width=0.83\textwidth]{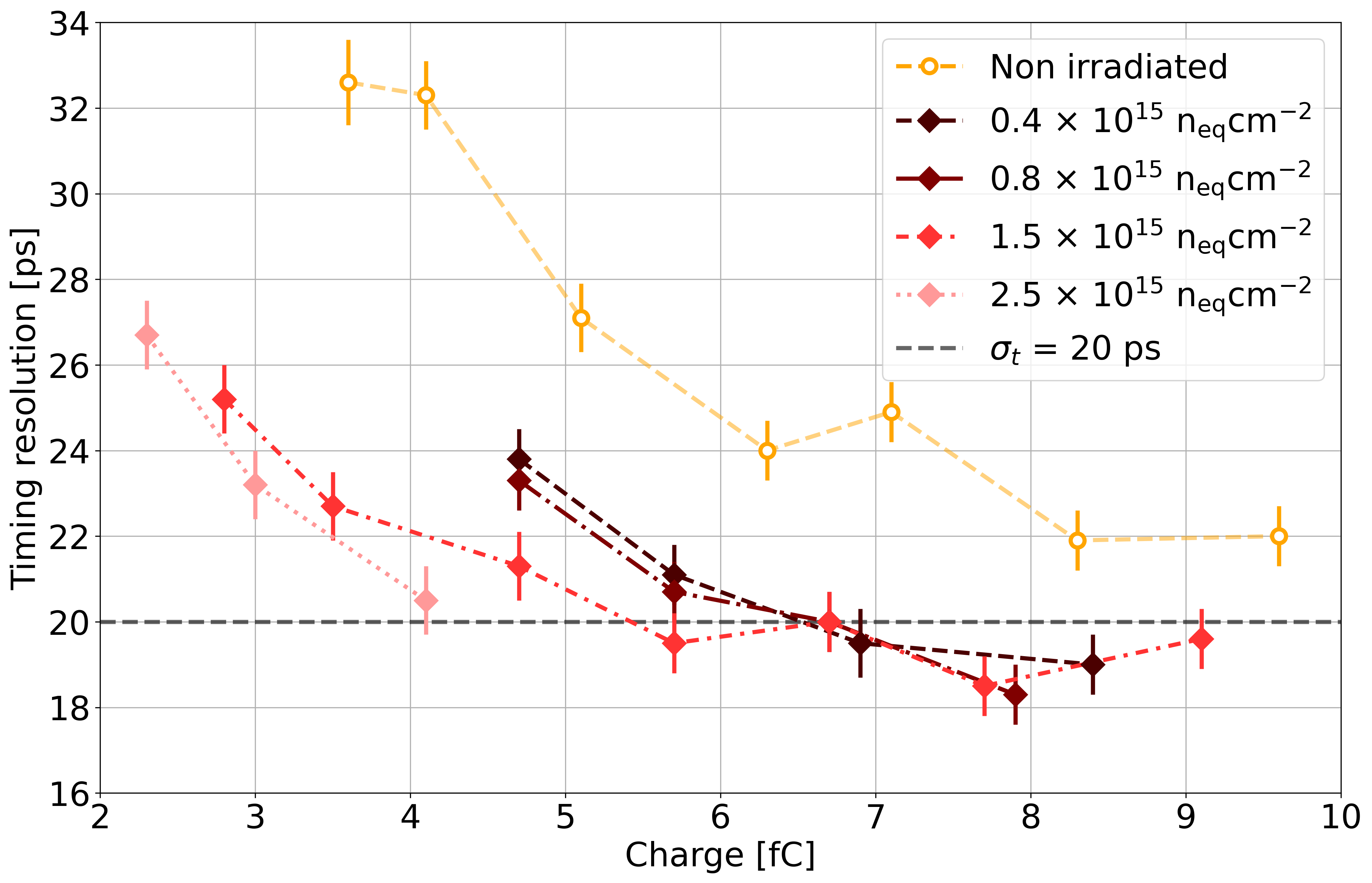}
\end{center}
\caption{The timing resolution of the irradiated 30~\micron-thick sensors as a function of collected charge, using data obtained between -50 and -35$^{\circ}$C. The non-irradiated sample tested at 18$^{\circ}$C is shown for reference.}
\label{fig:irrad_timeres_charge}
\end{figure}

In the irradiated samples, a timing resolution of 20~\ps is achieved across all irradiated sensors except for the 2.5~$\times$~10$^{15}$~\mevneut sample.
The results are shown in Figure~\ref{fig:irrad_timeres} as a function of the reverse bias, and in Figure~\ref{fig:irrad_timeres_charge} relative to the collected charge using data taken with irradiated samples at temperatures within the range of -50 and -35$^{\circ}$C.
The non-irradiated 30~\micron-thick sample is shown for reference.
The uncertainties in the \trdut estimates range between 0.7 and 0.8~\ps across the irradiated samples.
The results of the 2.5~$\times$~10$^{15}$~\mevneut indicate that a \trdut of 20.5~\ps is the best achievable before entering the SEB region~\cite{Ferrero:2025xby}.
The change in temperature does not affect the \trdut measurements, and hence no bias correction is applied.

\section{Conclusion}
\label{sec:summary}

The results of a test on a particle beam using 4~GeV/c electrons at the Test Beam Facility at DESY show that very thin sensors from the EXFLU production with substrate thicknesses below 35~\micron can achieve sub-25~\ps timing precision, and as good as 16.6~\ps in 20~\micron-thick sensors.
A two-plane LGAD tracker setup using the 20~\micron-thick samples achieved a timing resolution of 12.2~\ps.
The improvement with substrate thickness is understood as the signal \risetime and \landau are inherently smaller for thinner LGADs.

The minimum amount of charge required to achieve a given timing precision is observed to decrease linearly with active thickness. To achieve a sub-30~\ps timing resolution, the minimum collected charge is more than a factor of 2.5 smaller between the 35~\micron and 20~\micron-thick sensors, decreasing from $\sim$~6~\fC to $\sim$~2~\fC.

A study using irradiated 30~\micron-thick sensors demonstrated that a timing resolution of 18.3~\ps is achieved for sensors irradiated at fluences up to 1.5~$\times$~10$^{15}$~\mevneut, and 20.5~\ps at 2.5~$\times$~10$^{15}$~\mevneut before reaching the single-event burnout limit.

\section*{Acknowledgements}

\noindent This project has received funding from the European Union's Horizon 2020 research and innovation programme under Grant Agreements Nos 101004761 (AIDAInnova) and 101057511 (EURO-LABS), the European Union - Next Generation EU, Mission 4 Component 1 CUP D53D23002870001 (ComonSens), and the INFN CSN5 through the `eXFlu' research project. The measurements leading to these results have been performed at the Test Beam Facility at DESY, Hamburg (Germany), a member of the Helmholtz Association (HGF).

\bibliographystyle{elsarticle-num} 
\bibliography{cas-refs.bib}

\end{document}